%% file: NaID_SNgal.tex
\documentclass{aa}  

\usepackage{graphicx}
\usepackage[colorlinks=true,citecolor=blue]{hyperref}
\usepackage{txfonts}
\usepackage{newtxtext, newtxmath}
\usepackage[T1]{fontenc}
\usepackage{amsmath}
\usepackage{amssymb}
\usepackage{color}
\usepackage{xcolor}
\usepackage{hyperref}
\usepackage{pdflscape}
\usepackage{multirow}
\usepackage{threeparttable}
\usepackage{makecell}
\usepackage{bm}
\usepackage{arydshln}
\usepackage{float} 
\usepackage{multirow}

\newcommand{\Msun}{M$_\odot$}

\newcommand{\Zsun}{Z$_\odot$}

\newcommand{\kms}{\,km\,s$^{-1}$} 
\newcommand{\naid}{\ion{Na}{i}\,D\,}

\newcommand{\caii}{\ion{Ca}{ii}\,}
\newcommand{\ki}{\ion{K}{i}\,}

\usepackage{hyperref}
\definecolor{yaleblue}{rgb}{0.1,0.3,0.9}
\definecolor{lava}{rgb}{0.81, 0.06, 0.13}
\definecolor{forestgreen}{rgb}{0.0, 0.45, 0.13}
\hypersetup{colorlinks=true, linkcolor=lava, urlcolor=forestgreen, citecolor=yaleblue}

\newcommand{\orcid}[1]{\href{https://orcid.org/#1}{\includegraphics[width=10pt]{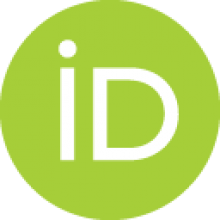}}}

\defcitealias{GG24}{Paper I}
\defcitealias{GG25}{Paper II}

\begin{document} 

\title{Narrow absorption lines from intervening material in supernovae}

\subtitle{III. Supernovae and their environments}

\author{
Claudia P. Guti\'errez
\inst{1,2}\orcid{0000-0003-2375-2064}
\and
Santiago Gonz\'alez-Gait\'an
\inst{3,4}\orcid{0000-0001-9541-0317}
\and
Joseph P. Anderson
\inst{3}\orcid{0000-0003-0227-3451}
\and
Llu\'is Galbany
\inst{2,1}\orcid{0000-0002-1296-6887}
}

\institute{
Institut d'Estudis Espacials de Catalunya (IEEC), Edifici RDIT, Campus UPC, 08860 Castelldefels (Barcelona), Spain\\
\email{cgutierrez@ice.csic.es}
\and
Institute of Space Sciences (ICE, CSIC), Campus UAB, Carrer de Can Magrans, s/n, E-08193 Barcelona, Spain
\and
European Southern Observatory, Alonso de C\'ordova 3107, Casilla 19, Santiago, Chile
\and
Instituto de Astrof\'isica e Ci\^encias do Espaço, Faculdade de Ci\^encias, Universidade de Lisboa, Ed. C8, Campo Grande, 1749-016 Lisbon, Portugal\\
\email{gongsale@gmail.com}
}
\date{}

 
\abstract
{
Narrow interstellar absorption features in supernova (SN) spectra serve as valuable diagnostics for probing dust extinction and the presence of circumstellar or interstellar material. In this third paper in a series, we investigate how the strength of narrow interstellar absorption lines in low-resolution spectra varies with SN type and host galaxy properties, both on local and global scales. Using a dataset of over 10000 spectra from $\sim1800$ low-redshift SNe, we find that Type Ia SNe (SNe~Ia) in passive galaxies exhibit significantly weaker narrow absorption features compared to CC-SNe and SNe~Ia in star-forming hosts (SNe~Ia-SF), suggesting lower interstellar gas content in quiescent environments. Within the star-forming hosts, the \naid\ equivalent-width distribution of SNe~II is much lower than that of both SNe Ia-SF and stripped-envelope SNe (SE-SNe). This result is somewhat unexpected, since CC-SNe are generally associated with star-forming regions and occur deeper within galactic disks, where stronger line-of-sight extinction would be anticipated. This suggests that the observed behaviour cannot be explained solely by absorption from the integrated interstellar medium (ISM) along the line of sight. Instead, if part of the absorption arises from material near the explosion, the similarity between the \naid\ EW distributions of SNe~Ia-SF and SE-SNe implies that comparable absorption signatures can emerge from distinct progenitor pathways. Possible explanations include (a) circumstellar material (CSM) expelled by the progenitor system before explosion, or (b) interaction of SN radiation with nearby patchy ISM clouds.
Our results highlight the diagnostic power of interstellar absorption features in revealing the diverse environments and progenitor pathways of SNe.
}

\authorrunning{Guti\'errez, Gonz\'alez-Gait\'an et al.}
\titlerunning{SN narrow lines and their environmental properties}
\maketitle
%

\section{Introduction}

The final stages of stellar evolution are occasionally marked by violent and bright explosions known as supernovae (SNe). These events display a wide range of observable characteristics, such as duration, brightness, colour, and spectral features, among others, that vary significantly across different SN types. One of the fundamental questions in SN science has always been what kind of progenitor star system gives rise to each distinct SN subtype. 

There are various methods for inferring the characteristics of progenitor stars from observations. Direct detections of progenitor stars in pre-explosion images of very nearby SNe offer the most direct constraints on the exploding stellar system, as long as one confirms that the progenitor really disappeared after the explosion \citep[e.g.][]{VanDyk02, VanDyk03, Mattila08, Smartt09, Folatelli16, VanDyk17}. Alternatively, comparisons between observed multi-wavelength light-curves and theoretical models \citep[e.g.][]{Bersten11, Bersten12, Morozova15, Kozyreva17, Martinez22, Moriya23, Gutierrez20, Gutierrez21, Gutierrez22}, as well as spectral modelling \citep[e.g.][]{Mazzali05, Jerkstrand12, Dessart13, Hillier19, Vogl19, Dessart23, Dessart24}, offer a powerful indirect route. These approaches, however, rely on high data quality and complex theoretical simulations. 

Another way to constrain the SN progenitors is by studying the environments of the SNe. For events with short delay times -- the period between star formation and explosion -- the surrounding environment can serve as a reliable proxy for progenitor properties.  Although this method is statistical and subject to various caveats  (e.g., spatial resolution, projection effects), it has proven powerful when applied across large samples \citep{Anderson15}. Those environmental studies commonly explore properties such as star formation rate \citep[e.g.][]{Anderson12, Galbany14, Habergham14, Kuncarayakti18} and metallicity \citep[e.g.][]{Arcavi10, Kuncarayakti13, Galbany16, Pessi23}, among others.

Supernovae are broadly classified into main categories based on their assumed progenitors: Type Ia SNe (SNe~Ia) originate from low-mass stars that have evolved into white dwarfs and accrete matter from a companion until thermonuclear runaway. In contrast, core-collapse (CC) SNe arise from massive stars that exhaust their nuclear fuel and cannot sustain fusion in their cores any longer. SNe~Ia are associated with progenitors of a large range of delay times, including long ones, while CC-SN progenitors are linked exclusively to short-lived progenitors. Within the CC sample, various subtypes include hydrogen-rich SNe (SNe~II), hydrogen-poor/deficient SNe (stripped-envelope SNe; SE-SNe) and interacting SNe (SNe-int), which show clear signatures of circumstellar material (CSM) interaction in their spectra.      

Understanding the environments in which different SN types occur is therefore key to constraining their progenitors. While the stellar evolution pathway determines the final stages of the star, explosion mechanisms, and the resulting SN, the surrounding interstellar medium (ISM) carries complementary information about the conditions in which the progenitors were born and evolved, as well as how the SN light interacts with intervening material. Its distribution and abundance can help trace the age and, with it, the initial mass of SN progenitors with short delay times. In particular, the narrow absorption lines from ISM species, such as sodium (\naid) in SN spectra, can unveil important clues about the SN progenitors.  

Narrow absorption lines are frequently observed in SN spectra \citep{Sternberg11, Maguire13, Phillips13, Gutierrez16, Clark21} and are often used as proxies for extinction \citep[e.g.][]{Turatto03, Phillips13}. While their use as direct extinction indicators comes with significant caveats, including line saturation, intrinsic scatter, contamination from CSM, multiple absorbing components and fake evolution of the line strength from broad ejecta line interference, these features provide valuable information as indirect tracers of both the progenitor’s birth environment and the physical conditions in its immediate surroundings. Consequently, the study of narrow absorption lines offers a powerful means of linking the intrinsic properties of SN progenitors with the extrinsic characteristics of their environments.

In \citet[][hereafter Paper I]{GG24}, we introduced a new robust technique to measure the pseudo-equivalent width (EW) and velocity (VEL) of interstellar lines in SN spectra, and investigated their evolution throughout the SN lifetime, finding no statistically significant change in those lines. In \citet[][hereafter Paper II]{GG25}, we extended the analysis by correlating these narrow features with local ($< 0.5$ kpc) and global host galaxy properties, validating their use as ISM tracers across a variety of environments. In this work, we further explore the connection between progenitor systems and their environments by dividing the sample into SN subtypes. We investigate how interstellar line measurements vary among these groups, aiming to reveal differences in local ISM content and potential clues about the nature of the exploding stars.

The paper is organised as follows. Section~\ref{sec:data} describes the data, measurements and methodology. The analysis and results are presented in Section~\ref{sec:res} and discussed in Section~\ref{sec:disc}. We conclude in Section~\ref{sec:conc}.

\section{Data, measurements and methodology}
\label{sec:data}

This section briefly summarises the data and methods used in this study. A complete description of the data and SN measurements can be found in \citetalias{GG24}, while the galaxy measurements and the methods used to analyse the data are explained in detail in \citetalias{GG25}.

\subsection{SN data and measurements}

The data used in this analysis consist of $>10000$ spectra for around 1800 low-redshift ($z\lesssim0.2$) SNe of different types: type Ia SNe (SNe Ia), type II SNe (SNe II), stripped-envelope SNe (SE-SNe) and interacting SNe (SNe-int). We measure the pseudo-equivalent widths (EWs) and velocities (VELs) of the narrow interstellar absorption lines for each SN spectrum. When multiple spectra for a given SN are available, we stack the flux-to-continuum and measure the EW, VEL and their uncertainties. We use individual spectra with a high enough signal-to-noise ratio (S/N $>15$) and with a shallow continuum slope. Table~\ref{table:cuts} summarises the quality cuts established in \citepalias{GG24}, which were designed to minimise biases in the EW measurements. Further details can be found in \citetalias{GG24}.

\begin{table}
\centering
\caption{Spectral quality cuts.}
\label{table:cuts}
\begin{tabular}{ccc}
\hline
\hline
Parameter       &           Cut          \\ 
\hline
Redshift        &       $z>0.004<0.2$    \\
Signal-to-noise &        S/N$>15$        \\
Continuum slope & $m_c<0.002~$\AA$^{-1}$ \\
\hline
\end{tabular}
\end{table}

\subsection{Galaxy measurements}

Host galaxy photometry was obtained with \textsc{hostphot}\footnote{\url{https://hostphot.readthedocs.io/en/latest/}} \citep{Muller-Bravo22}, and the resulting measurements were used to derive stellar population properties through \textsc{prospector}\footnote{\url{https://prospect.readthedocs.io/en/latest/}} \citep{Johnson21-prospector, Leja17} fits to both local (circular apertures of 0.5 kpc radius centred on the SN) and global photometry. From these fits, we obtained parameters such as stellar mass ($M_*$), stellar metallicity ($Z_*$), age ($t_{\mathrm{age}}$), e-folding time ($\tau$), attenuation ($A_V$), dust index ($n$), star formation rate (SFR) and specific SFR (sSFR). Additional host galaxy information was gathered from the NASA/IPAC Extragalactic Database (NED) database\footnote{\url{https://ned.ipac.caltech.edu/}} and the Asiago catalogue\footnote{\url{http://graspa.oapd.inaf.it/asnc/}}, including SN angular offset ($\Delta\alpha$), SN normalised offset ($\overline{\Delta\alpha}$), SN directional offset ($\Delta\alpha_{\mathrm{DLR}}$), galaxy inclination ($i$), galaxy type (T-type). Further details of the methodology and tools employed can be found in \citetalias{GG25}.

\section{Analysis and Results}
\label{sec:res}

Given that SNe~Ia are found in star-forming (SF) and passive (pass) galaxies, we split this sample into two groups: SNe~Ia-SF and SNe~Ia-pass. Specifically, we define SNe~Ia-pass as those occurring in galaxies with T-type$<=0$, while SNe~Ia-SF correspond to events in galaxies with T-type$>0$. Therefore, for the following analysis, we consider five SN classes: SNe~II, SE-SNe, SNe-int, SNe~Ia-SF and SNe~Ia-pass. Applying the methodology presented in \citetalias{GG25}, we compare the EW and VEL distributions of the narrow interstellar \naid\ line among different samples by generating cumulative distributions. We use the Kolmogorov-Smirnov (K-S) test \citep{ks-test1,ks-test2} to evaluate the statistical significance of our results. For increased robustness, we do a "$z$-matched bootstrap K-S test" (see \citetalias{GG25}), in which 1000 random bootstrap sub-samples are generated from the original distributions, making sure that the redshift distributions of the two samples are consistent with each other. We
quote the median p-value of all these iterations and the probability of the p-value being less than 0.05 ($P_{MC}$). We also show the non-parametric Spearman's rank coefficient, $r_s$, between two variables \citep{Spearman1904}, when appropriate. In Section~\ref{gal-prop}, we first show the different local and global properties divided by SN type to compare with previous studies, while in Section~\ref{ewna}, we present the EW and VEL of \naid\, divided according to SN type. In Section~\ref{sec:other-lines}, we investigate the EW and VEL of other narrow lines in SNe, whereas in Section~\ref{sec:na-sntype}, we compare the EW and VEL of the lines of different SN types with various galaxy properties.

\subsection{Galaxy properties per SN type}
\label{gal-prop}

\input{table_ks_gen}

K-S statistics, including the $p$-value and the probability $P_{MC}$ of the $p$-value being lower than 0.05 from a bootstrap analysis \citepalias[see][]{GG25} obtained from comparisons of different galaxy properties divided according to SN type are presented in Table~\ref{table:KSgen}. As shown in this table, the most significant differences arise when comparing different SN types with SNe Ia-pass.

\begin{figure}
\centering
\includegraphics[width=\columnwidth]{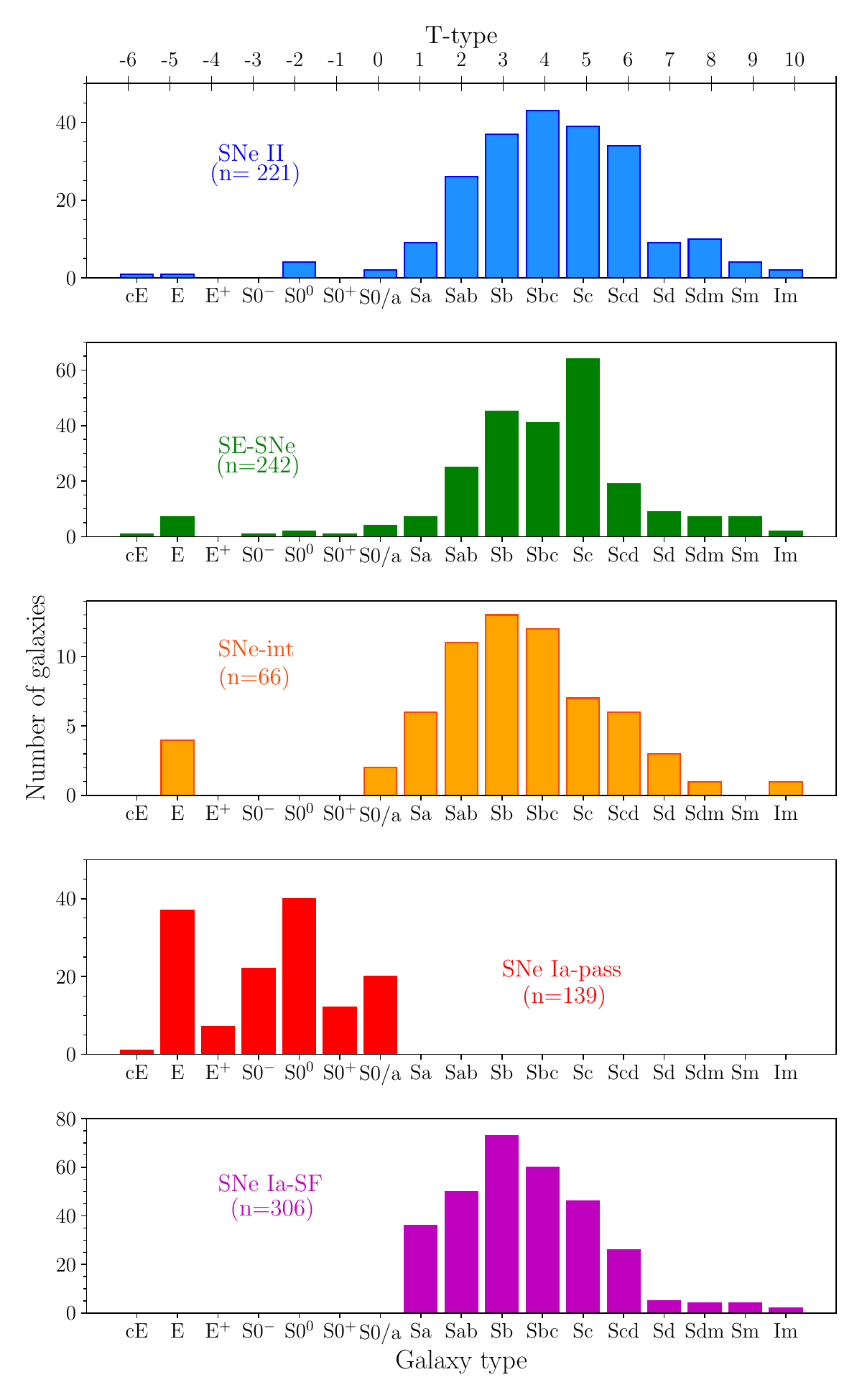}
\caption{Galaxy type distribution of our sample split by the SN type. The SN type and the number of events are presented at the top of each panel.
}
\label{fig:galtype}
\end{figure}

To further illustrate these differences, Figure~\ref{fig:galtype} shows the distribution of host galaxy morphological types for each SN subtype in our targeted historical sample. As expected, CC-SNe are found in star-forming galaxies, while SNe~Ia occur in both passive and star-forming hosts. For most SN types, the distribution of galaxy types approximately follows a normal distribution, centred around intermediate spiral types; the only exception is observed for SNe~Ia-pass, which exhibit a bimodal distribution: one peak corresponding to E galaxies and the other peak to S0$^0$ galaxies, but this comes from the fact that the E galaxy type is often given as a generic elliptical class.

Interestingly, we also detect CC-SNe occurring in early-type (elliptical) galaxies (T-type$<=0$), where such events are uncommon. Specifically, our sample includes 8 SNe~II, 16 SE, and 6 Int in elliptical hosts. The occurrence of CC-SNe in early-type galaxies has been noted in previous studies \citep[e.g.][]{Sedgwick21, Irani22} and is potentially explained by either residual or recent star formation or, in some cases, by misclassifications of host galaxy morphology \citep{Gomes16}.
In this analysis, we find that while uncommon, CC-SNe may occur in early-type galaxies at a non-negligible number. This is consistent with galaxy studies, which show that in a modest fraction, typically a few tens of per cent, of nearby early-type galaxies exhibit detectable levels of star formation \citep[e.g.][]{Schawinski07, Young09}.

In terms of offset and inclination, SNe~Ia-pass are significantly different from the other types, which is expected following their definition: early-type galaxies are morphologically distinct, being larger, more ellipsoidal in shape, and lacking clear disks. So it is unsurprising that SNe~Ia-pass typically occur at larger separations from the galaxy centre and seemingly with lower inclinations than the other SN types. 

The clear difference between SNe~Ia-pass and the rest of SN types, which occur mostly in SF galaxies, is also evident in global host properties such as the stellar mass, the specific star formation (sSFR$_0^G$), and the stellar age ($t_{\mathrm{age}}^G$): SNe~Ia-pass occur in more massive, older and less star-forming galaxies. Interestingly, we find that SNe~Ia-SF also preferentially occur in more massive galaxies with respect to CC-SNe, e.g. their mass distributions strongly differ from those of SNe~II  ($P_{MC}=93\%$) and SE~SNe ($P_{MC}=88\%$).

Since the global attenuation ($A_V^G$) is influenced by the stellar mass, meaning that more massive star-forming galaxies have more dust, leading to higher attenuation \citep{Garn10,Zahid13,Duarte23}, we normalised the $A_V^G$ by the stellar mass (M$_*^G$), to isolate this effect. This new parameter, which we refer to as specific attenuation ($sA_V^G=A_V^G$/M$_*^G$), allows for a mass-independent comparison (see Appendix~\ref{ap:sAv}). After normalisation, we find that SNe~II and SE-SNe occur in galaxies with higher attenuation per unit mass, followed by SNe~Ia-SF, whereas SNe~Ia-pass hosts have the lowest $sA_V^G$ values.

\begin{figure*}
\centering
\includegraphics[width=0.49\textwidth]{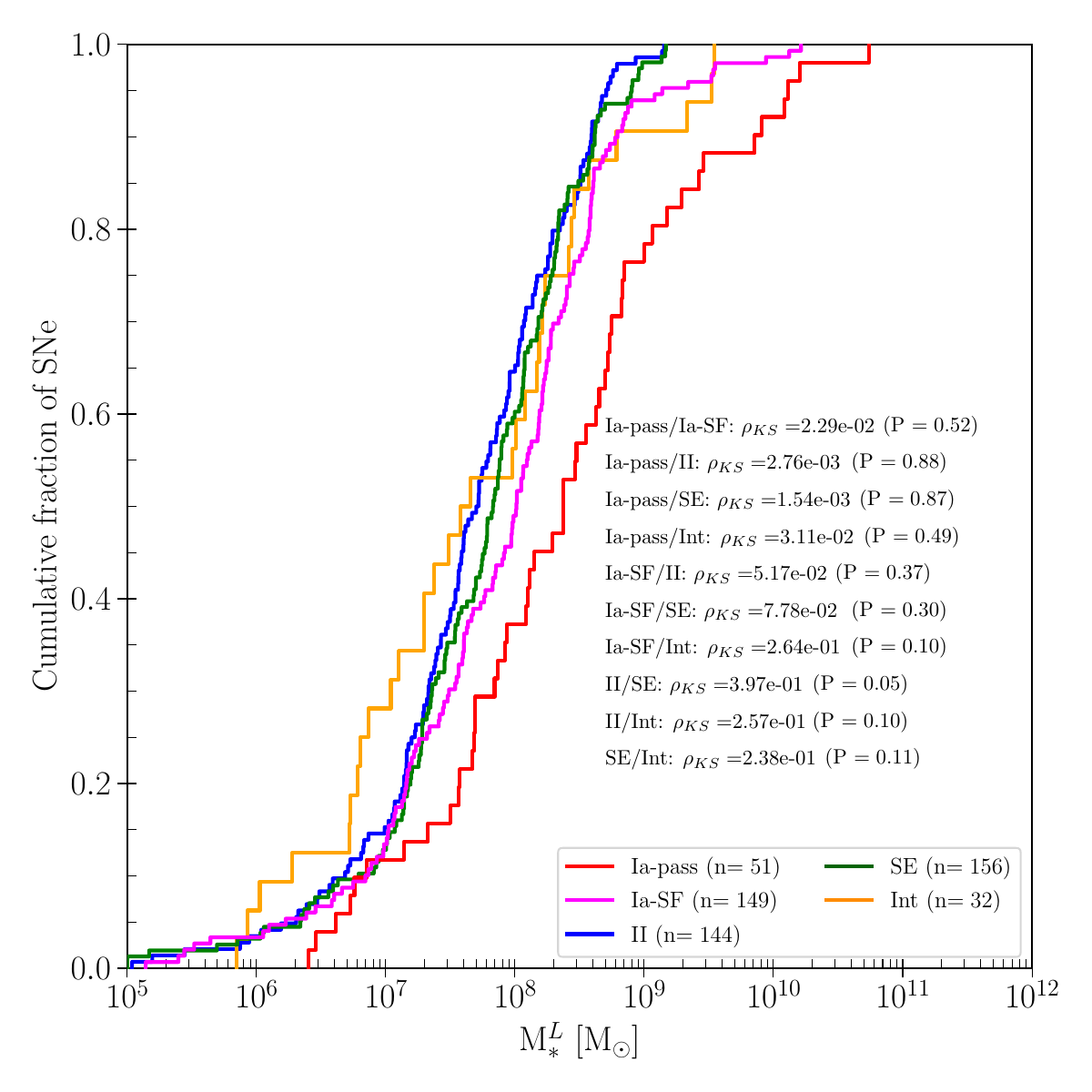}
\includegraphics[width=0.49\textwidth]{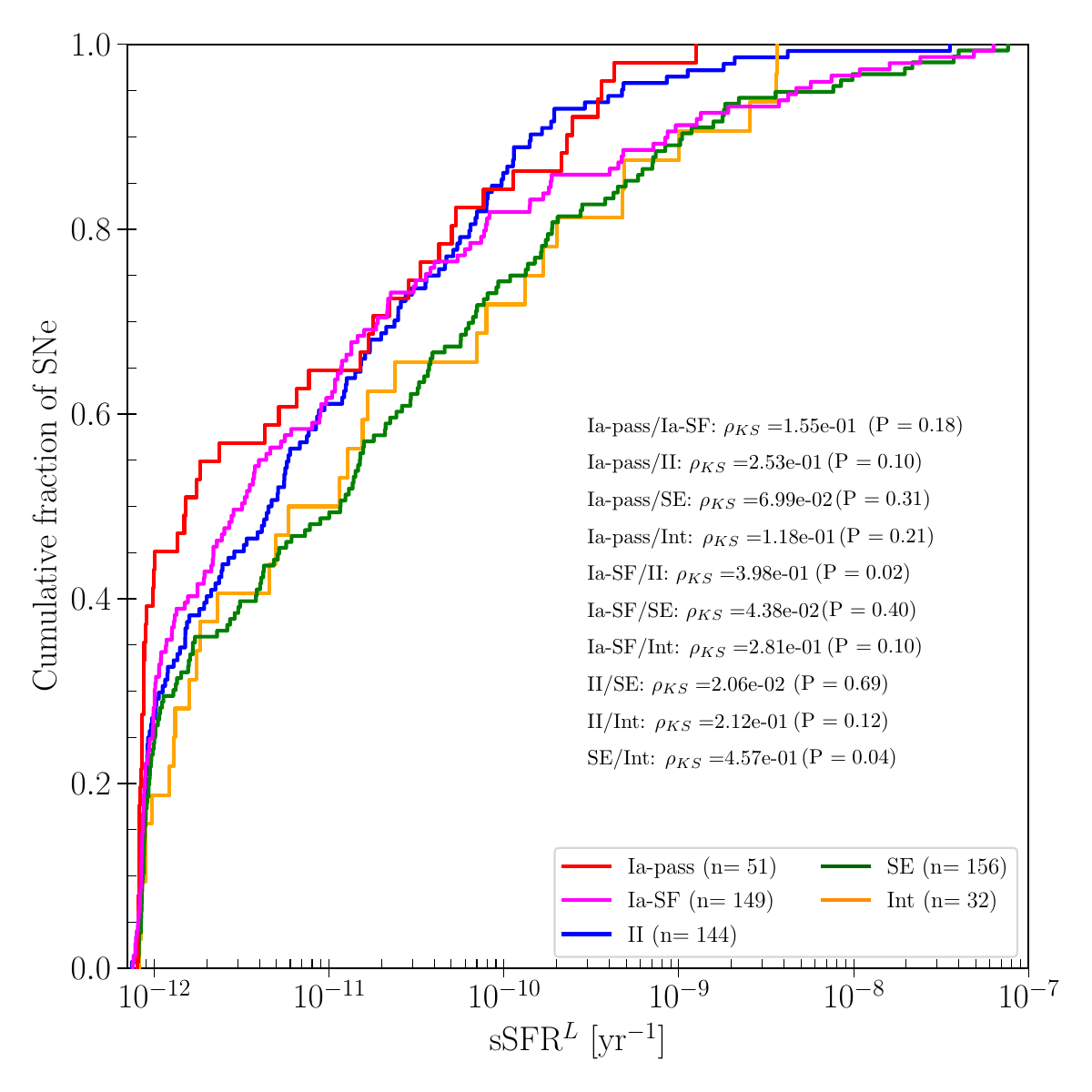}
\includegraphics[width=0.49\textwidth]{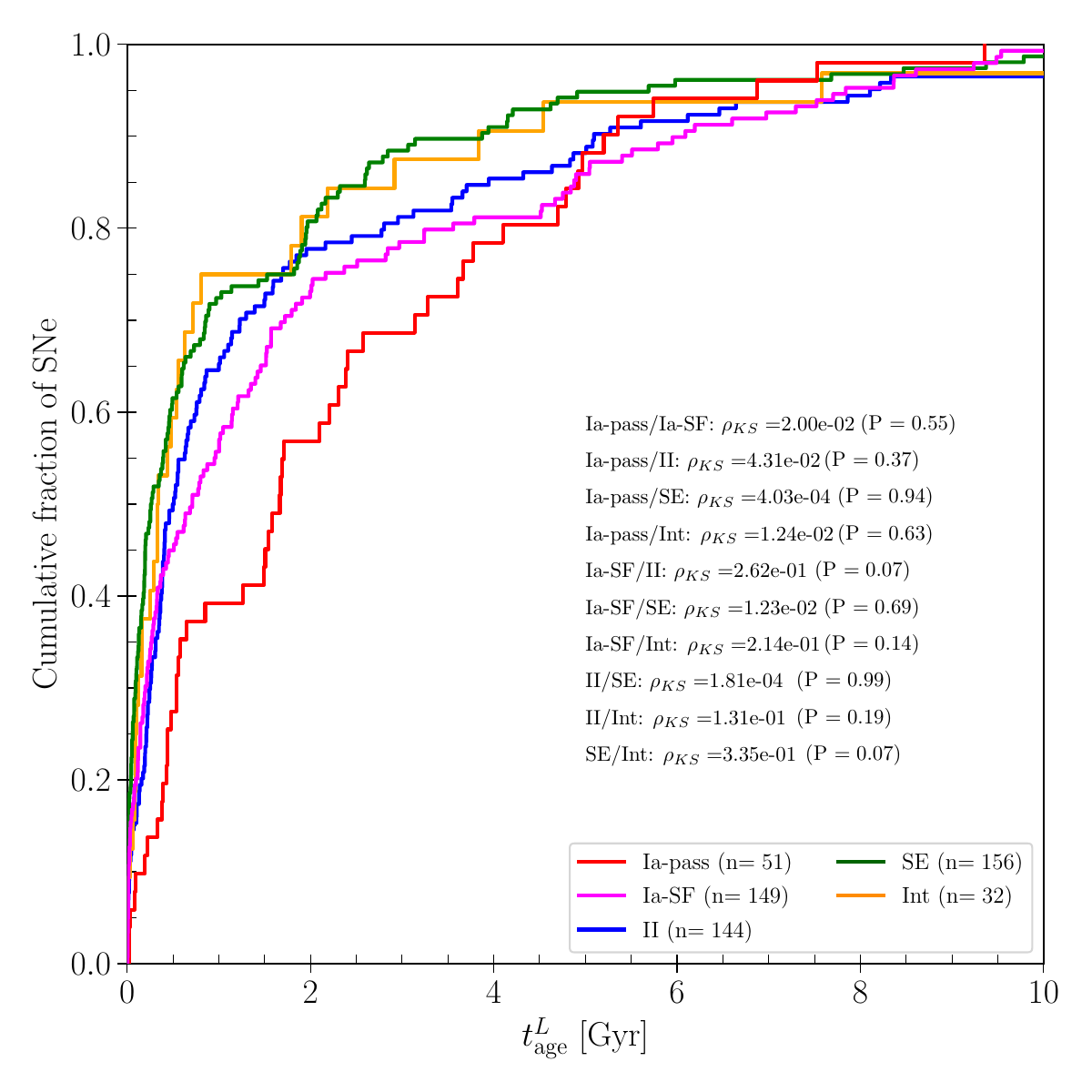}
\includegraphics[width=0.49\textwidth]{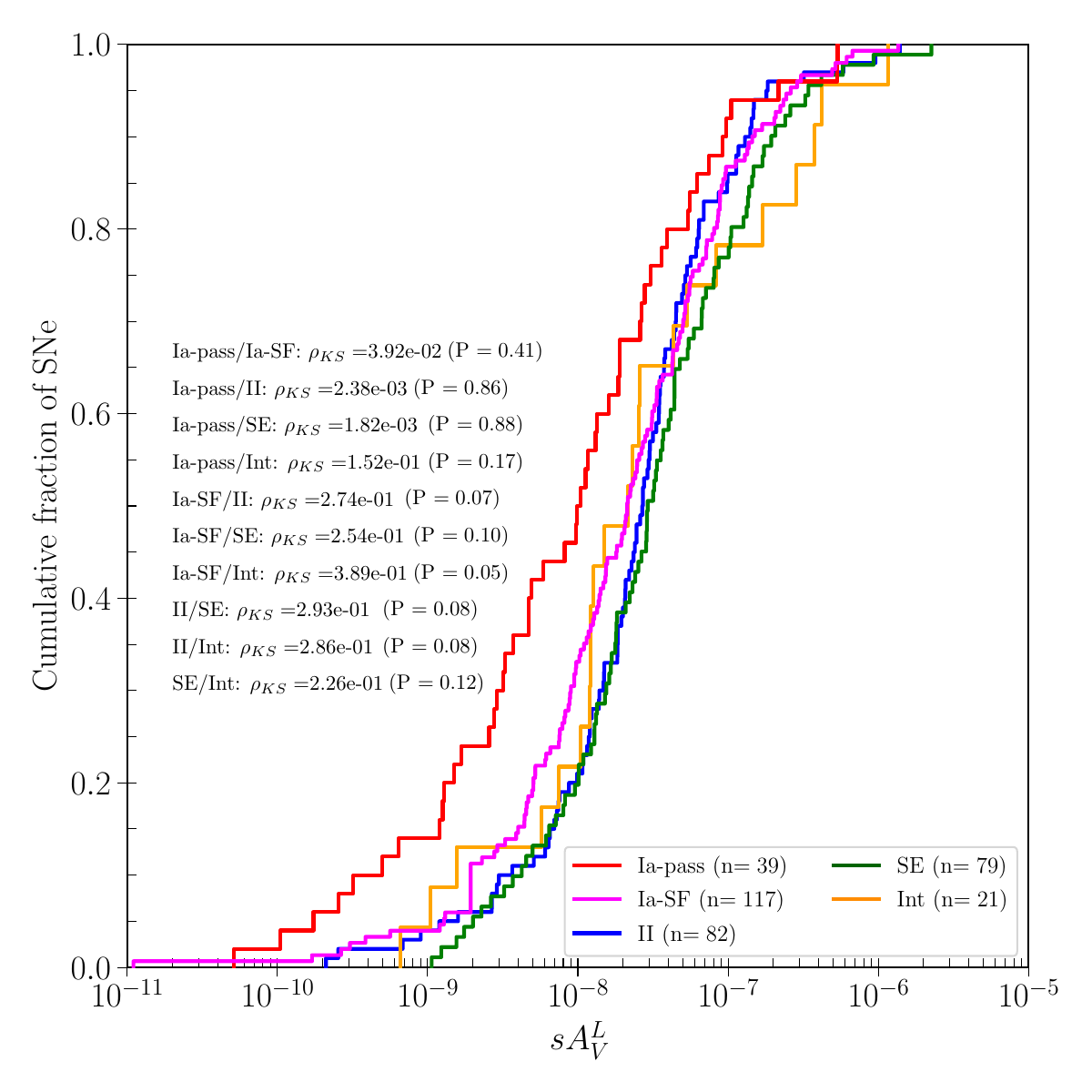}
\caption{Cumulative distributions of the local M$_*^L$, sSFR$^L$, $t_{\mathrm{age}}^L$ and $sA_V^L$ ($A_V^L/M_*^L$) for different SN types. The number of events of each SN type and the results from the K-S test comparing two different populations are also shown in each panel.
}
\label{fig:prop_type}
\end{figure*}

Moving on to local host galaxy properties, we find that the local stellar mass (M$_*^L$) follows the same overall trend observed globally, though with lower statistical significance: SNe~Ia-pass preferentially explode in more massive local environments, followed by SNe~Ia-SF, while SNe~II and SE typically occur in lower mass regions (see upper left Figure~\ref{fig:prop_type}). This is consistent with the expectation that older progenitors follow the stellar mass more closely. 

For the local sSFR (sSFR$_0^L$), significant differences appear primarily between SNe~II and SE ($P_{MC}=69\%$), indicating that SE-SNe occur in regions with higher star formation (see upper right Figure~\ref{fig:prop_type}).
SNe~II and SNe~Ia-SF have, in fact, consistent sSFR$_0^L$ distributions, suggesting no substantial local differences in star formation activity between them. 

Regarding the local stellar age ($t_{\mathrm{age}}^L$), the difference between SNe~Ia-pass and other SN types identified at the global scale becomes even more pronounced locally: $P_{MC}=94\%$ with SE-SNe, $P_{MC}=55\%$ with SNe~Ia-SF and $P_{MC}=63\%$ with SNe-int. A very strong difference is also detected between SNe~II and SE-SNe ($P_{MC}=99\%$). These findings indicate that SNe~Ia-pass preferentially occur in older local stellar environments, followed by SNe~Ia-SF and SNe~II.  In contrast, SE-SNe and SNe-int tend to explode in regions with younger stellar populations (lower left Figure~\ref{fig:prop_type}). 

When comparing  the local specific attenuation ($sA_V^L$; see lower right Figure~\ref{fig:prop_type}), we find that SNe~Ia-pass displays the lowest $sA_V^L$ values, with strong differences observed between SNe~Ia-pass and both SE-SNe ($P_{MC}=88\%$) and SNe~II ($P_{MC}=86\%$). CC-SNe and SNe~Ia-SF seem to all have similar specific attenuations, whereas SNe~Ia-pass environments are comparatively less dusty.

\subsection{EW and VEL distributions for \naid}
\label{ewna}

\begin{figure*}
\centering
\includegraphics[width=\columnwidth]{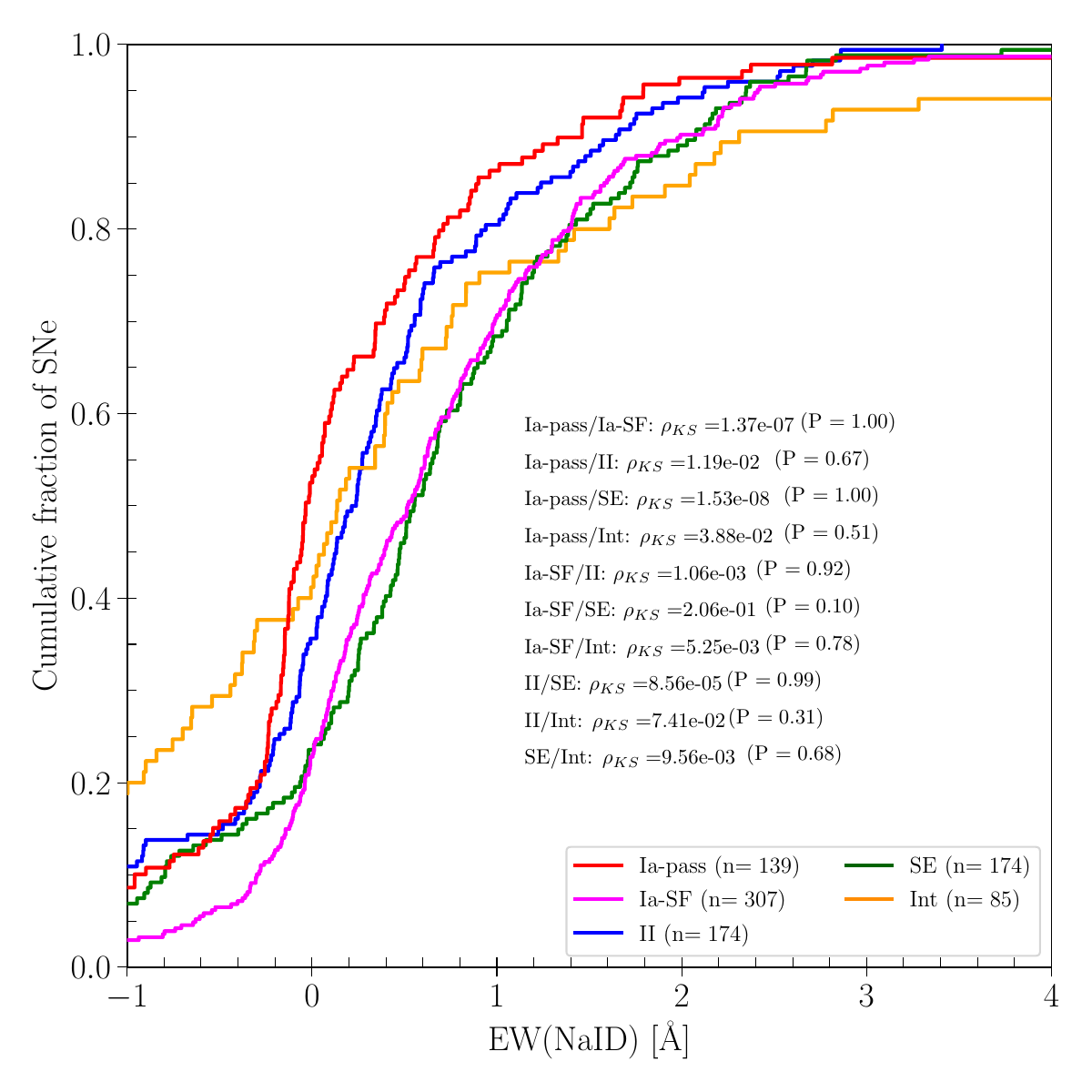}
\includegraphics[width=\columnwidth]{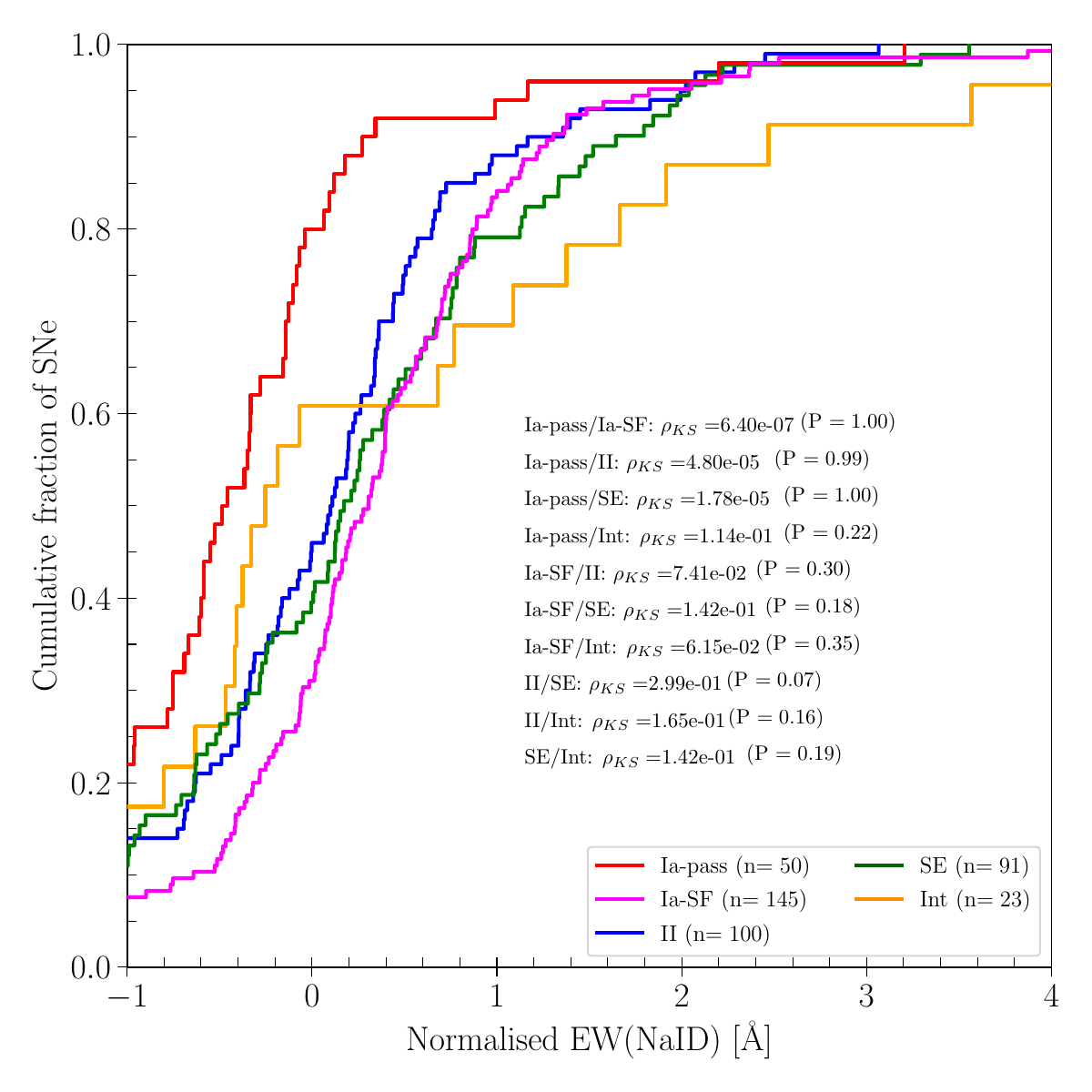}
\caption{Cumulative distributions of the measured \naid\ EW \textbf{(left)} and the EW normalised by the offset and M$_*^L$ (\textbf{right)} for each SN type. The number of events of each SN type and the results from the K-S test comparing two different populations are also shown in each panel.
}
\label{fig:dist_full}
\end{figure*}

\input{table_median}

In this section, we explore differences in EW and VEL based only on the SN type. The left panel of Figure~\ref{fig:dist_full} shows the measured \naid\ EW distribution of our sample. Significant variations in EW are observed across different SN types. Specifically, SNe~Ia-pass exhibit the lowest EW values with a median of $\overline{EW}=-0.03$ \AA\ (see Table~\ref{table:median}), followed by SNe-int (0.15 \AA), SNe~II (0.23 \AA), SNe~Ia-SF (0.52 \AA) and SE-SNe (0.55 \AA). K-S tests comparing these distributions reveal that SNe~Ia-pass differ significantly from most other SN types, with the null hypothesis of shared parent population rejected at a confidence level exceeding 50\%. This trend aligns with the fact that passive galaxies, where SNe~Ia-pass occur, have substantially lower gas content \citep{Young11}.

An interesting result is the seemingly low median value of \naid\ EWs measured for SNe-int. Taking the mean, however, results in a much higher value of 0.47 \AA, and from Figure~\ref{fig:dist_full} (left panel) we see that the distribution is first full of low values but rapidly increases to very high values, possibly indicating two populations. Although these SNe interact with circumstellar material and have traditionally been associated with very massive progenitors, environmental studies  \citep[e.g.,][]{Habergham14} have shown that SNe-int exhibit a relatively weak association with ongoing star formation, as traced by H$\alpha$ emission (but see top right panel of Figure~\ref{fig:prop_type} and Table~\ref{table:KSgen}). This weak association could help explain the SNe-int with low \naid\ EWs. At the same time, the presence of several high EW values among SNe-int suggests that a subset of these objects exhibits significant absorption. Notably, SNe-int also show a higher frequency of negative measured EWs (mostly consistent with zero) compared to other SN types. To test for the impact of this effect, we exclude all SNe with EW$+\sigma_{\mathrm{EW}}<0$, and repeat the analysis. After applying this cut, the results remain unchanged, and no significant differences emerge. This suggests that the observed trend is likely intrinsic rather than an artefact of measurement uncertainties.

A more surprising result from Figure~\ref{fig:dist_full} (left panel) is that SNe~Ia-SF have a significantly stronger EW distribution than SNe~II and more similar to what is observed for SE-SNe. The EW of SNe~Ia-SF and SE-SNe are in fact statistically consistent with being drawn from the same parent distribution. The discrepancy with SNe~II is in agreement with the trends identified in Section~\ref{gal-prop}, which show that SNe~II are in older and less star-forming environments than SE SNe. However, SNe~Ia-SF have properties more in line with SNe~II (similar local sSFR and slightly older local age) but have EW of \naid\, more similar to SE~SNe. At first glance, one might suspect that this result is driven by redshift effects, given that SNe~Ia are intrinsically brighter and their \naid\ features may be easier to detect at larger distances. However, as we are using a redshift-matched bootstrap K-S test \citepalias[see Section~\ref{sec:res} and][for more details]{GG25} to ensure that the redshift distributions of the two samples are directly comparable, these results are not driven by redshift effects  (see Appendix~\ref{ap:redshift}). Instead, it indicates that the result is robust and likely reflects a genuine physical trend.

To mitigate the known dependence of the strength of the \naid\ narrow line with galaxy properties that, in turn, also correlate with SN types, we normalised the EW measurements by $\overline{\Delta\alpha}$ and M$_*^L$, following the analytic correlations found in \citetalias{GG25} (see Appendix~\ref{ap:nEW}). Therein, we showed that the \naid\ EW correlates moderately with $\overline{\Delta\alpha}$ ($r_s^{\mathrm{EW}}=-0.44$), SFR$_0^L$ ($r_s^{\mathrm{EW}}=0.39$), $A_V^L$ ($r_s^{\mathrm{EW}}=0.35$) and M$_*^L$ ($r_s^{\mathrm{EW}}=0.32$). Subtracting the normalised offset ($\overline{\Delta\alpha}$) relation removes this dependence, but since the SFR$_0^L$ and $A_V^L$ also depend on the M$_*^L$, we additionally subtract the M$_*^L$ relation. The results of this "normalised EW" are shown in the right panel of Figure~\ref{fig:dist_full}. Although visually the cumulative distributions look more separated, the difference according to the K-S statistics for the majority of the comparisons decreased (see Appendix~\ref{ap:downsampling}). This is partly due to the decrease in the number of events, as not all SNe have local galaxy parameters measured. The resulting distributions in CC-SNe and SNe~Ia-SF become more similar; however, the distribution of SNe Ia-pass differs from these, exhibiting much lower EW values. Statistically, the comparison between SNe Ia-pass and SNe II is the only one that increases, showing now a stronger difference, going from $P_{MC}=67\%$ to $P_{MC}=99\%$. For SNe~Ia-pass and SNe-int, the difference observed before the normalisation disappears, declining from $P_{MC}=51\%$ to $P_{MC}=22\%$. Visually, the distribution of SNe-int shows more clearly the bi-modality between no and strong \naid absorption.

\begin{figure}
\centering
\includegraphics[width=\columnwidth]{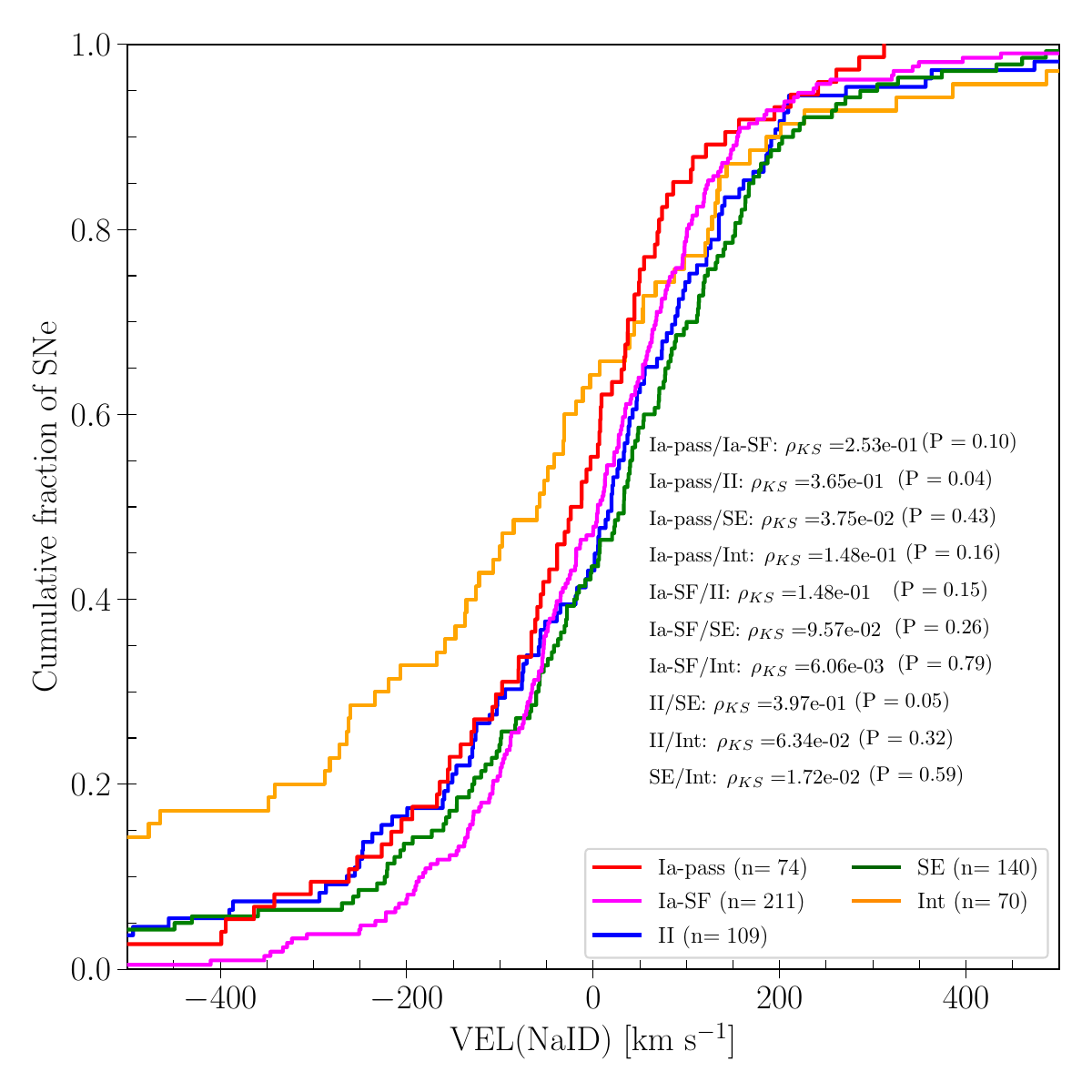}
\caption{Cumulative distributions of the \naid\ VEL divided into the SN type. The number of events of each SN type and the results from the K-S test comparing two different populations are also shown in each panel.
}
\label{fig:dist_vel}
\end{figure}

Figure~\ref{fig:dist_vel} shows the \naid\ VEL distributions for each SN type. Since the velocities are impossible to measure when the lines are inexistent or very weak, we do not consider objects with |EW|$<0.3$\AA. We observe strong differences for SNe-int compared to SNe~Ia-SF ($P_{MC}=79\%$) and SE-SNe ($P_{MC}=59\%$). For the rest of the comparisons, no significant differences were found. Overall, SNe-int exhibit the highest fraction of SNe with blueshifted absorption with a median velocity of $\overline{VEL}=-59$ \kms. The remaining SNe types have a median between $-18$ and $33$ \kms. This finding suggests that the VEL may be largely independent of the SN type. The stronger blueshifted VEL in the SNe-int sample could be linked to CSM moving towards the observer. However, confirming this interpretation remains challenging due to the large measured uncertainties and the limited number of available observations. We therefore carefully investigated potential measurement biases, in particular contamination from nearby narrow emission or absorption features. A visual inspection of the spectra and the relevant spectral lines supports the reliability of the measured values. In addition, restricting the sample to only positive absorption values (EW$>0.3$ \AA) produces similar results. Similarly, limiting the sample to measurements with small relative uncertainties (EW/ERR$>3$) also provides similar median negative blueshifted lines, indicating that the observed trend is robust against these selection criteria. Alternative explanations beyond a CSM origin may involve nearby dust clouds accelerated toward the observer by SN radiation pressure \citep{Hoang19}. However, the energy released by SNe-int is comparable to that of SNe~Ia, for which no corresponding excess blueshift is observed. This suggests that radiation-driven dust acceleration alone is unlikely to account for the observed blueshift in the SNe-int sample. Table~\ref{table:median} provides the median, average and median absolute deviation (MAD) for the \naid\ VEL.

\subsection{EW and VEL distributions for other narrow lines}\label{sec:other-lines}

We also measured the EW and VEL of \caii H, \caii K,  \ki 1, \ki 2, DIB-4428, DIB-5780 and DIB-6283. When comparing across different SNe subtypes, we find statistically significant differences for \caii H, \caii K, DIB-4428, DIB-6283 and \ki 2, as follows: In general, SNe~Ia-pass tend to show the weakest absorption (i.e. the lowest EWs), while either SE-SN or SNe~Ia-SF show the strongest. This is in agreement with our findings for the \naid\ EW (Section~\ref{ewna}). Specifically, for \caii H, the most notable differences occur when comparing SNe~Ia with other subtypes. The statistical significance of the differences between SNe~Ia-pass and other types ranges from 58\% to 72\%. These differences become more pronounced when comparing SNe~Ia-SF to CC-SNe, with significance levels between 70\% and 90\%. In this case, SNe~Ia-SF exhibit stronger \caii H EW absorption, while SNe-Ia-pass again show the weakest. 

For \caii K, the EW distributions also reveal strong differences. SNe~Ia-pass differ significantly from SE-SNe, SNe~II and SNe~Ia-SF ($P_{MC}=94\% - 100\%$). A moderate difference is also found between SNe~Ia-SF and SE-SNe ($P_{MC}=68\%$). SE-SNe show the highest EW values for this line, while SNe~Ia-pass continue to exhibit the lowest values. 

When comparing the EW distribution of DIBs, we find notable differences in DIB-6283 and DIB-4458. For DIB-6283, significant differences are observed between SE-SNe and both SNe~Ia-pass ($P_{MC}=80\%$), SNe~Ia-SF ($P_{MC}=78\%$), as well as between SNe-int and both SNe~Ia-SF ($P_{MC}=95\%$) and SNe~II ($P_{MC}=66\%$). SNe-int shows the strongest DIB-6283 EW, followed by SE-SNe. For DIB-4458, significant differences are only found between SNe~Ia-SF and both SNe-int ($P_{MC}=96\%$) and SNe-II ($P_{MC}=83\%$), with SNe~Ia-SF showing the highest EW values. Finally, for the \ki 2, we observe strong differences between SNe-Ia-SF and both SNe~II ($P_{MC}=85\%$) and SE-SNe ($P_{MC}=62\%$).

Regarding the VEL distributions, differences are only found for the \caii lines. For \caii K, these differences are seen between SNe-int and SNe~Ia-SF ($P_{MC}=79\%$), SE-SNe ($P_{MC}=75\%$) and SNe~Ia-pass ($P_{MC}=57\%$). For \caii H, we find a difference between SNe~Ia-SF and SNe-int ($P_{MC}=66\%$). In all cases, SNe-int consistently show the bluest VEL values among the compared groups. No significant VEL differences are detected among SN subtypes for the other lines.
Tables~\ref{table:KSgen-dib} and \ref{table:median-dib} (Appendix~\ref{ap:tables}) summarise the statistical properties of the narrow absorption lines discussed in this section: Table~\ref{table:KSgen-dib} reports the K-S statistics for the EW distributions across SN subtypes, while Table~\ref{table:median-dib} provides the median, average and MAD values.

\subsection{\naid\ differences based on the SN type and galaxy properties}
\label{sec:na-sntype}

In \citetalias{GG25}, we showed that the SN EW traces several global and local galaxy properties such as the T-type, the inclination, the galactocentric offset, the local mass, sSFR, age and dust attenuation. Here, we investigate how these properties correlate with our gas indicators when divided by SN type. The K-S statistics and correlations for \naid\ EW/VEL distributions for each SN type, divided according to galaxy properties, are presented in Tables~\ref{table:KS-Ia-pass} --  \ref{table:KS-int} (Appendix~\ref{ap:tables}). The properties with strong K-S test statistics are highlighted in bold, according to their $z$-matched bootstrap probability of being drawn from different parent populations, i.e. probability higher than 50\% of a K-S p-value less than 0.05: $P_{MC}(\rho<0.05) > 50\%$. 

SNe~Ia-SF, II, SE and Int with stronger \naid\ EWs are generally located in galaxies with higher inclinations, as shown in \citetalias{GG25}, although there are no significant statistical differences when comparing their inclination-split \naid\ EW/VEL distributions. Regarding the offsets from the galaxy centres, the trend identified in \citetalias{GG25} persists when separating by SN types. Across all classifications, SNe at lower and higher offsets from the galaxy centre exhibit different \naid\ EW distributions. For normalised offsets, the difference is statistically significant for most SN types ($P_{MC}>60$\%), with weaker \naid\ lines found at larger distances from the galaxy centre. This trend is especially pronounced for SNe~Ia-SF ($P_{MC}=100$\%). For the angular and DLR offsets, only SNe~Ia (both Ia-pass and Ia-SF) show significant differences in their \naid\ EW distributions ($P_{MC}>80$\%). While a similar trend is seen for CC-SNe, the statistical significance is lower. In contrast, we find no significant differences in the VEL distributions across inclination or offset for any SN type.

Regarding the global properties of the SN host galaxies, as for the entire SN sample, we do not find significant differences in the \naid\ EW/VEL distributions. However, the trends observed for local environmental properties do change significantly. The SN types showing the most statistically significant differences are SNe~Ia-SF and SE-SNe. For SNe~Ia-SF, we find strong differences ($P_{MC}\sim58-99\%$) in the EW distributions for various parameters, including recent (< 100 Myr) SFR$_0^L$ ($P_{MC}=99\%$), sSFR$_0^L$ ($P_{MC}=78\%$), $A_V^L$ ($P_{MC}=91\%$), M$_*^L$ ($P_{MC}=84\%$) and $t_{\mathrm{age}}$ ($P_{MC}=58\%$). This indicates that SNe~Ia-SF with higher \naid EW values tend to be located in more massive, actively star-forming, dustier (more attenuated) and younger environments. For SE-SNe, the most significant differences are found with $A_V$ ($P_{MC}=100\%$), SFR$_0^L$ ($P_{MC}=91\%$), sSFR$_0^L$ ($P_{MC}=80\%$), M$_*^L$ ($P_{MC}=86\%$) and the attenuation slope $n$ ($P_{MC}=78\%$), suggesting that SE-SNe with higher \naid\ EW are associated with dustier, more massive and actively star-forming galaxies. Although no statistically significant differences are found for the other SN types (SNe~Ia-pass, II, Int), similar trends are seen: higher \naid\ EWs are typically associated with more massive, star-forming, younger and dustier environments.  
This reinforces the conclusions from \citetalias{GG25}, suggesting that the SN \naid\ narrow lines better trace the local environment rather than the global properties of the host galaxies.

\section{Discussion}
\label{sec:disc}

Using over 10000 spectra from $\sim1800$ low-redshift SNe of different types (SNe~II, SE-SNe, SNe-int, SNe~Ia-SF and SNe~Ia-pass) along with their corresponding host galaxy information, we have investigated the properties of the host galaxies and how they affect the strength of narrow interstellar absorption lines (especially \naid) in SN spectra. In this section, we interpret the key findings presented above and place them in the context of previous studies.

\subsection{Comparison to previous SN environmental studies}

Studies focusing on SN host galaxies are common and often employ a variety of samples and analytical techniques, offering more detailed insights into the environments of different SN types \citep[see][and references therein]{Anderson15}. 

Using the pixel statistics technique (NCR), \citet{Anderson12} found that SNe~Ic show the strongest association with H$\alpha$ emission (tracing very young stellar populations), followed by SNe~Ib, SNe~II and finally SNe~Ia, which shows the weakest degree of association \citep[also see,][]{Anderson08, Anderson15}. These results suggest that SNe~Ic more closely trace the SF of their host galaxies than SNe~II. \cite{Galbany14} reported similar results when comparing SNe~Ia, II and Ibc, but using Integral Field Spectroscopy (IFS) of the hosts. Our analysis supports these previous conclusions; in particular, we find that the sSFR$_0^L$ of SE-SNe are systematically higher than those from SNe~II and SNe~Ia (see top-right panel of Figure~\ref{fig:prop_type} and Table~\ref{table:KSgen}). 

When interacting SNe are included in these NCR studies, the results have been unexpected, suggesting that at least some SNe-int may arise from older populations and lower-mass progenitors \citep{Habergham14, Anderson15, Kuncarayakti18}. Despite the limited number of SNe-int in our sample, we observe this trend only at the low-end of the sSFR$_0^L$ distribution, rapidly increasing to high values, more similar to SE-SNe at the high-end. This could mean that SNe-int may have a mix of progenitors \citep[also see,][]{Galbany18}, although the statistics are too low to draw any firm conclusions.

We reach similar results when analysing the local $t_{\mathrm{age}}^L$. As expected, SNe~Ia (both Ia-pass and Ia-SF) are associated with older populations, consistent with long progenitor lifetimes. Among CC-SNe, we find a strong difference in the $t_{\mathrm{age}}$ distributions between SNe~II and SE-SNe ($P_{MC}=99\%$), providing robust evidence that SE-SNe preferentially occur in significantly younger environments. Interestingly, SNe-int appear to arise from progenitors with intermediate ages, suggesting that they may occupy an evolutionary phase between those of SE-SNe and SNe~II.

We also compare our stellar masses, M$_*$, with previous studies. Given that metallicity is a key parameter in stellar evolution, and considering that metallicity (\Zsun) estimates obtained from \textsc{prospector} are not fully reliable (see, e.g. \citealt{Duarte23}), we adopt the M$_*$ as a proxy for metallicity. From a theoretical perspective, metallicity-driven mass loss in single-star progenitors predicts that SE-SNe should preferentially occur in more massive, metal-rich environments \citep[e.g.][]{Meynet05, Eldridge08}.

Our sample covers a wide range of masses, from relatively low-mass galaxies with log(M$_*^G)\sim8$ \Msun\ to more massive hosts reaching log(M$_*^G)\sim12$ \Msun. This mass range is comparable to the host galaxy sample analysed by \citet{Galbany18}. Studies including a narrower mass range  \citep[e.g.][log(M$_*^G)\sim9-10$ \Msun]{Galbany14} reported systematic differences across all SN types, with SNe~Ia at higher M$_*$ and SE-SNe in the lower end. In our data, SNe~II and SE-SNe exhibit similar M$_*$ distributions (with SNe~II tending toward slightly lower masses), both much lower than those of SNe Ia (Ia-SF and Ia-pass). This trend is consistent at both global and local scales. While higher metallicity should, in principle, favour SE-SNe in more massive environments, the inclusion of Type Ic broad-line SNe (SNe~Ic-BL), known to occur preferentially at lower metallicities, can broaden and shift the M$_*$ distribution, reducing the contrast with SNe~II. Nevertheless, the number of SNe~Ic-BL in our sample is very small, and excluding them does not significantly alter the overall distributions. Therefore, the observed similarity in M$_*$ distributions between SNe~II and SE-SNe appears to be robust, indicating that metallicity is not the dominant parameter shaping the CC-SN diversity \citep[also see][]{Anderson12}. Instead, other parameters, such as binarity, stellar rotation or zero-age main sequence (ZAMS) mass, are likely to play a more significant role in determining the final SN subtype. However, to definitively disentangle these effects, direct metallicity indicators (instead of proxies like stellar mass) for each CC-SN subtype (SNe~II, IIb, Ib, Ic, Ic-BL) across large, unbiased samples are essential. While our sample is somewhat biased against the faintest galaxies, our findings highlight the importance of including low-mass hosts to fully capture the diversity of SN progenitor environments.

\subsection{Comparison to previous studies analysing narrow lines}

Statistical studies focusing on the narrow interstellar lines in SN spectra remain limited. In SNe~Ia, these narrow lines have primarily been analysed to investigate and constrain progenitor scenarios \citep[e.g.][]{Sternberg13, Maguire13, Phillips13, Clark21}. However, analyses directly linking the strength of these narrow lines to galaxy properties are rare. To our knowledge, only two such studies have been published. The first, by \citet{Anderson15}, explored how the \naid\ absorption in SNe~II changes with environment and found that events with detectable \naid\ tend to occur closer to the centres of their host galaxies. Our results confirm this trend: SNe~II located at smaller offsets exhibit higher EW values, as do all other SN types. A K-S test comparing the EW distributions shows a statistically significant difference, with $P_{MC}=86\%$ (see Table~\ref{table:KS-II}). 

The second work, presented by \citet{Nugent23}, analysed a sample of SNe~Ia with different ejecta velocities and explored correlation with host galaxy properties. They found no statistically significant difference in the \naid\ distributions between high- and low-velocity SNe~Ia, but noted that high velocity events tend to exhibit higher EWs and are more concentrated near the centre of their host galaxies. Although our current analysis does not split SN~Ia based on their expansion velocities, we find that SNe~Ia-SF generally have larger \naid\ EWs and are more located in central regions compared to SNe~Ia-pass. We also find that, when dividing the EW measurements by the median of the normalised offset, both SNe~Ia-pass and SNe~Ia-SF with larger EWs are more frequently located in the central regions of their hosts. The difference between central and outer-region events is statistically significant, with $P_{MC}=71\%$ for SNe~Ia-pass and $P_{MC}=99\%$ for SNe~Ia-SF. While these findings suggest a connection between line strength and local environment, the possible association between high-velocity SNe Ia and star-forming hosts remains inconclusive. We will address this question in more depth by examining SN Ia sub-populations in a forthcoming study.

\subsection{\naid\ EW as tracer of local environments}

In \citetalias{GG25}, we showed that the narrow absorption lines in SN spectra are effective tracers for statistically probing the ISM properties of host galaxies. Specifically, we found significant differences in \naid\ EW when splitting the sample by local M$_*^L$ and local sSFR$_0^L$, with higher EW values associated with more massive galaxies and actively SF regions. In this work, we extended that analysis by examining EW distributions per SN type, revealing that SNe~Ia-SF and SE-SNe drive the strongest differences. For both types, we identify five environmental parameters that show statistically significant differences, four of which are shared: SFR$_0^L$, A$_V^L$, M$_*^L$ and sSFR$_0^L$. Additionally, SNe~Ia-SF show a dependence on $t_{\mathrm{age}}$, which can be attributed to the large range of ages observed in SNe~Ia, as opposed to CC-SNe. SE-SNe also exhibit differences in the $n$ parameter, but this comes from the strong correlation between $A_V$ and $n$ \citep[see e.g.][]{Duarte23}: since SE-SNe are the SN type with the largest \naid\ differences with $A_V$ ($P_{MC}$ =100\%), it is not surprising that $n$ also shows strong p-values. Our results suggest that SNe~Ia-SF and SE-SNe with stronger \naid\ lines tend to occur in dustier, more massive and more actively star-forming environments. This reinforces the idea that the narrow \naid\ absorption line is a sensitive tracer of local ISM, providing insight into the immediate environments of SN progenitors for these two SN types.

To better understand why two SN classes originating from very different stellar populations, one typically from older (Ia-SF) and the other from younger (SE-SNe) progenitors, exhibit similar environmental trends with \naid\ EW, we investigate deeper into their local host galaxy conditions. For this analysis, we used a Monte Carlo and applied the Fasano-Franceschini (F-F) test \citep{Fasano87}, a non-parametric, multidimensional generalisation of the K-S test. Using the parameters $t_{\mathrm{age}}$ (since SNe~Ia-SF and SE-SNe originate from different progenitor populations) together with the local parameters with the strongest differences (M$_*^L$, sA$_V^L$ and sSFR$_0^L$)\footnote{Since SFR$_0^L$, A$_V^L$ show strong dependency on the M$_*^L$, we adopt the normalised quantities sA$_V^L$ and sSFR$_0^L$ to account for this effect.}, we find that the environments of SNe Ia-SF are actually statistically different from those of SE-SNe (p$_{MC}$-value$=0.009\pm0.042$; $P_{MC}$ =85\%). Specifically, SNe~Ia-SF tend to explode in more massive environments with lower current SF activity. If we remove M$_*^L$ from the environmental parameters and we repeat the F-F test, we still find significant environmental differences between SNe~Ia-SF and SE-SNe (p-value$=0.014\pm0.064$; $P_{MC}$ =77\%).

The apparent similarity in the \naid\ distributions of SNe~Ia-SF and SE-SNe (K-S test with $P_{MC}$ =10\% before normalisation; $P_{MC}$ =18\% after normalisation; Figure~\ref{fig:dist_full}) is in stark contrast with the clearly distinct progenitor environments found according to the multi-dimensional F-F test. This suggests that similar ISM absorption signatures can arise from different environments and perhaps different progenitor scenarios. These scenarios may involve binary interaction, progenitor mass loss, or processes related to the explosion itself. More specifically, the EW behaviour could originate from the presence of more nearby material around the explosion site, either (a) from CSM ejected by the progenitor system before the explosion, or (b) through interaction of the SN radiation with patchy ISM material in the immediate vicinity.

A fraction of SNe~Ia exhibit \naid\ absorption in excess of what would be expected based on their A$_V$ values inferred from light curves \citep{Phillips13, Maguire13}. One interpretation is that this additional absorption originates from CSM associated with the progenitor system. Alternatively, intense SN radiation can affect nearby dust clouds by inducing cloud-cloud collisions \citep{Hoang17} or imprinting centrifugal forces on dust grains through radiative torques, leading to their disruption \citep{Hoang19, Giang20}. Such processes can release metals previously locked in dust grains, increasing the fraction of free sodium atoms in the local environment. The strength of this effect depends on the intensity of the SN radiation and the proximity of the dust clouds. Since SNe~Ia are generally more energetic than CC-SNe, such scenarios could partially explain the higher abundance of sodium in SNe~Ia-SF without the need for CSM.

\section{Conclusions}
\label{sec:conc}

In this paper, we have analysed the narrow interstellar lines in SNe spectra and investigated their connection to both local and global properties of the host galaxies. Our analysis combines the EW and VEL measurements presented in \citetalias{GG24} for over 1800 SNe, with the host galaxy properties characterised in \citetalias{GG25}.
Our main results are:

\begin{itemize}

\item CC-SNe occur at a non-negligible number in early-type galaxies, in agreement with previous studies. Clear host differences emerge among SN types: SNe Ia-pass occur in older, more massive, and less dusty systems, while SNe Ia-SF are found in more massive but dustier hosts, resembling CC-SNe. Consistently, SNe~II and SE-SNe show the highest specific attenuation, both globally and locally.

\item SE-SNe display the highest local sSFR, reinforcing their association with more massive progenitors. In contrast, SNe-int occupy an intermediate regime in both age and sSFR, suggesting a mixed origin between SE-SNe and SNe II progenitor populations.

\item SNe~Ia in passive galaxies exhibit significantly weaker narrow absorption features compared to CC-SNe and SNe~Ia-SF, suggesting lower interstellar gas content in quiescent environments. 

\item The \naid\ EW distribution of SNe~II is much lower than that of both SNe Ia-SF and SE-SNe, which is unexpected given their association with SF regions and deeper galactic locations. This could indicate that absorption (for SESN and SNe~Ia-SF) may include more nearby material beyond the line-of-sight ISM.

\item The similarity between SNe Ia-SF and SE-SNe \naid\ EW distributions, despite their different environment, suggests that different progenitors and environments can produce comparable absorption, possibly through CSM expelled before explosion or SN interaction with patchy ISM clouds.

\item Across all measured narrow interstellar lines (\naid, \caii H, \caii K, DIB-4428, DIB-5780, DIB-6283, \ki 1 and \ki 2), SNe~Ia-pass consistently show the weakest EWs, while either SE-SN or SNe Ia-SF show the strongest.

\end{itemize}

Detailed analyses focused on specific SN types will be presented in forthcoming papers. These studies will particularly explore how SN spectral and photometric properties correlate with host galaxy characteristics and the strength of narrow interstellar absorption features. The goal is to disentangle the physical mechanisms behind the observed diversity among SN subtypes and to constrain progenitor scenarios better using environmental diagnostics.

\begin{acknowledgements}

We thank the anonymous referee for the comments and suggestions that have helped us to improve the paper.
C.P.G. acknowledges financial support from the Secretary of Universities and Research (Government of Catalonia) and by the Horizon 2020 Research and Innovation Programme of the European Union under the Marie Sk\l{}odowska-Curie and the Beatriu de Pin\'os 2021 BP 00168 programme. 
C.P.G. and L.G. recognise the support from the Spanish Ministerio de Ciencia e Innovaci\'on (MCIN) and the Agencia Estatal de Investigaci\'on (AEI) 10.13039/501100011033 under the PID2023-151307NB-I00 SNNEXT project, from Centro Superior de Investigaciones Cient\'ificas (CSIC) under the PIE project 20215AT016 and the program Unidad de Excelencia Mar\'ia de Maeztu CEX2020-001058-M, and from the Departament de Recerca i Universitats de la Generalitat de Catalunya through the 2021-SGR-01270 grant. We acknowledge the financial support from the Mar\'ia de Maeztu Thematic Core at ICE-CSIC. 
S.G.G thanks the ESO Scientific Visitor Programme. \\

This research has made use of the \textsc{python} packages \textsc{astropy} \citep{astropy:2013,astropy:2018,astropy:2022}, \textsc{numpy} \citep{numpy}, \textsc{matplotlib} \citep{matplotlib}, \textsc{scipy} \citep{scipy}, \textsc{pandas} \citep{pandas,mckinney-pandas}.

\end{acknowledgements}

%
\bibliographystyle{aa} 
\bibliography{Bibliography}


\appendix

\section{Specific attenuation}
\label{ap:sAv}

In this work, we introduce the term "specific attenuation" ($sA_V$) as the attenuation (A$_V$) normalised by stellar mass (M$_*$). Attenuation is strongly correlated with stellar mass because more massive galaxies generally contain more dust. This is due to a larger fraction of stars enriching the ISM with metals, which are needed to form dust grains, as well as the greater gravitational potential retaining the dust. As a result, these galaxies naturally exhibit higher attenuation values. Without normalisation, observed trends in A$_V$ could simply mirror underlying trends in M$_*$, which itself correlates with various galaxy properties such as metallicity, SFR, age, and even scattering \citep{Duarte25}. 

\begin{figure}[!h]
\centering
\includegraphics[width=\columnwidth]{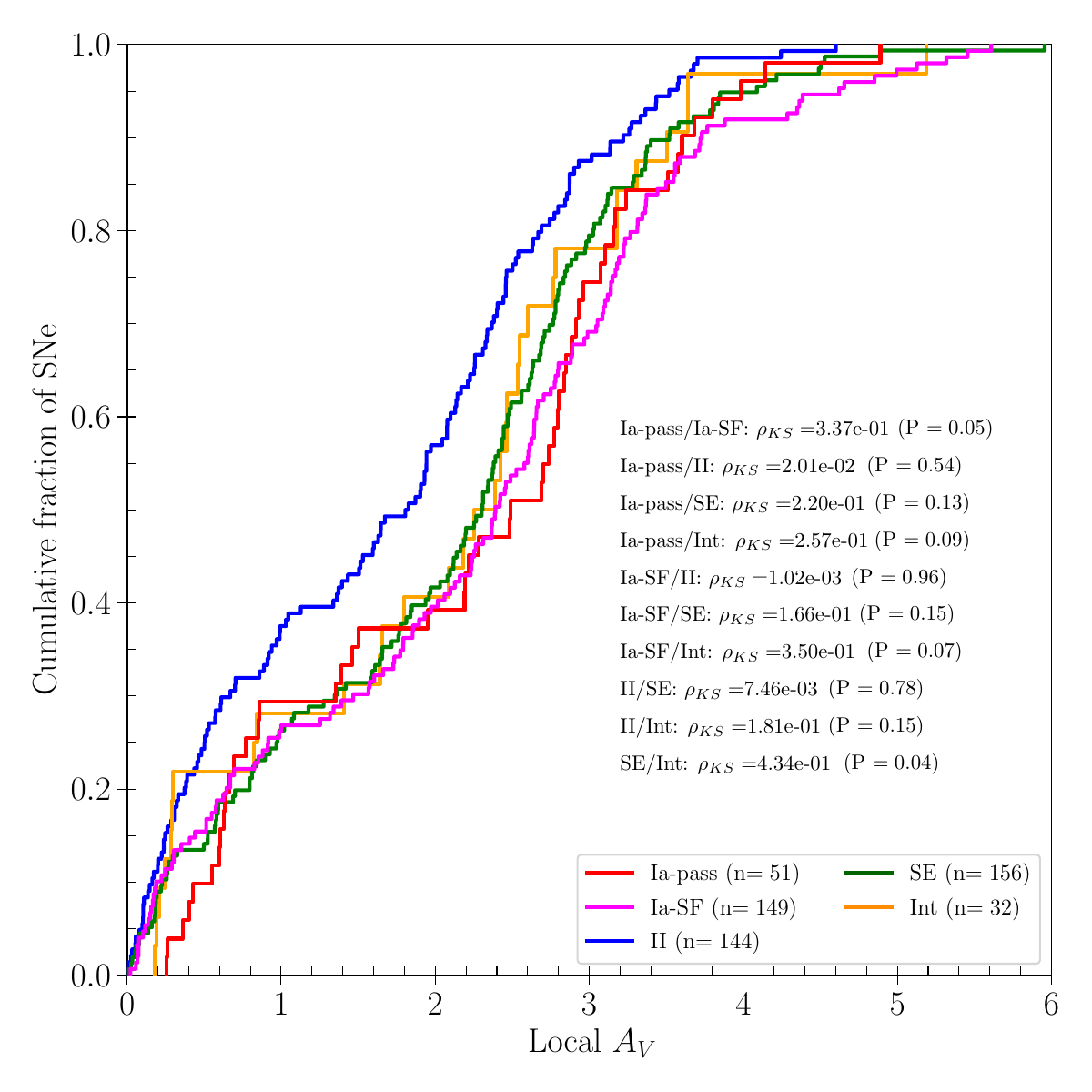}
\caption{Cumulative distributions of the local $A_V^L$ for different SN types. The number of events of each SN type and the results from the K-S test comparing two different populations are also shown in each panel.
}
\label{fig:avlocal}
\end{figure}

Figure~\ref{fig:avlocal} shows the $A_V$ distributions before normalisation, while the lower right panel of Figure~\ref{fig:prop_type} shows the normalised $A_V$ values (i.e. $sA_V$). This normalisation accounts for the intrinsic scaling between mass and dust content, allowing us to isolate dustiness relative to galaxy mass. By examining $sA_V$, we can better assess how dust properties vary across different SN host environments independently of galaxy mass. For example, a high $sA_V$ indicates a galaxy that is unusually dusty for its mass, which may reflect more active SF or denser ISM conditions. Conversely, low $sA_V$ values suggest relatively dust-poor environments, often associated with more quiescent galaxies. Thus, specific attenuation provides a more refined understanding of the dust environments where different SN progenitors arise.

\section{Redshift effects}
\label{ap:redshift}

\begin{figure}[!h]
\centering
\includegraphics[width=\columnwidth]{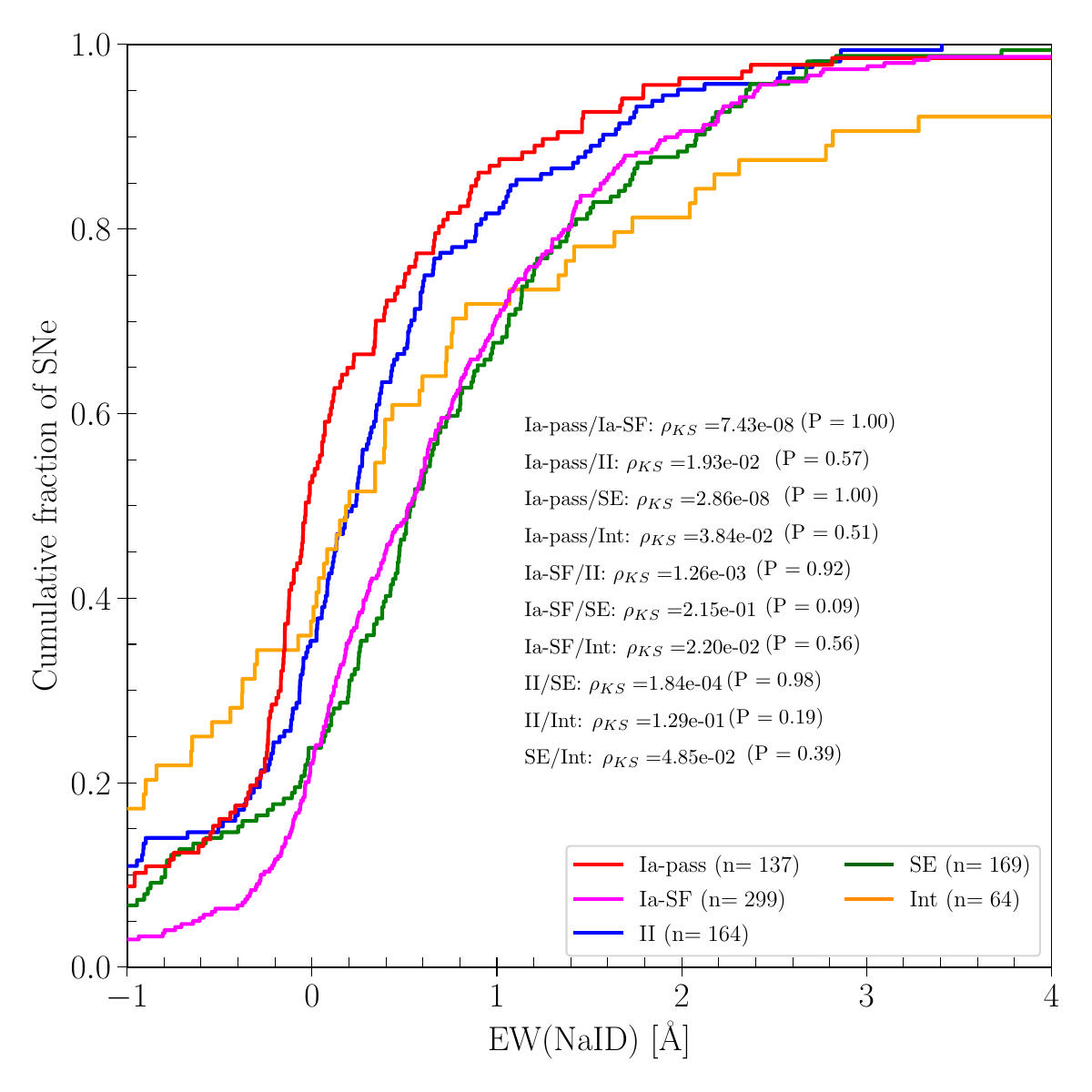}
\caption{Cumulative distributions of the measured \naid\ EW for a subsample of events restricted to $z<0.05$.
}
\label{fig:redshift}
\end{figure}

Figure~\ref{fig:redshift} shows the cumulative distributions of the measured \naid\ EW or each SN type (same as Figure~\ref{fig:dist_full}) but restricted to a subsample of events with $z<0.05$. A comparison between Figures~\ref{fig:dist_full} and \ref{fig:redshift} shows that the distributions remain largely unchanged, indicating that the results are not driven by redshift effects.

\section{Normalised \naid\ EW}
\label{ap:nEW}

In \citetalias{GG25}, we identified a dependence of the strength of the \naid\ narrow line on $\overline{\Delta\alpha}$, SFR$_0^L$ and M$_*^L$. For galaxy offset, \naid\ EW shows an exponential decline with increasing radius. To quantify this behaviour, we computed the rolling median of EW as a function of the normalised offset. We then removed this radial dependence by subtracting from each EW measurement the value predicted by the rolling median at its corresponding offset. This normalisation effectively isolates intrinsic EW variations that are independent of radial position within the galaxy. Normalising by offset removes nearly all correlations. However, because M$_*^L$ also influence SFR$_0^L$ and $A_V^L$, we applied an analogous correction using M$_*^L$. The resulting \naid\ EW distributions before and after the normalisations are shown in Figure~\ref{fig:dist_full}.

\section{Down-sampling affects}
\label{ap:downsampling}
\begin{figure*}
\centering
\includegraphics[width=0.67\columnwidth]{./Figures/dist_full_ew.pdf}
\includegraphics[width=0.67\columnwidth]{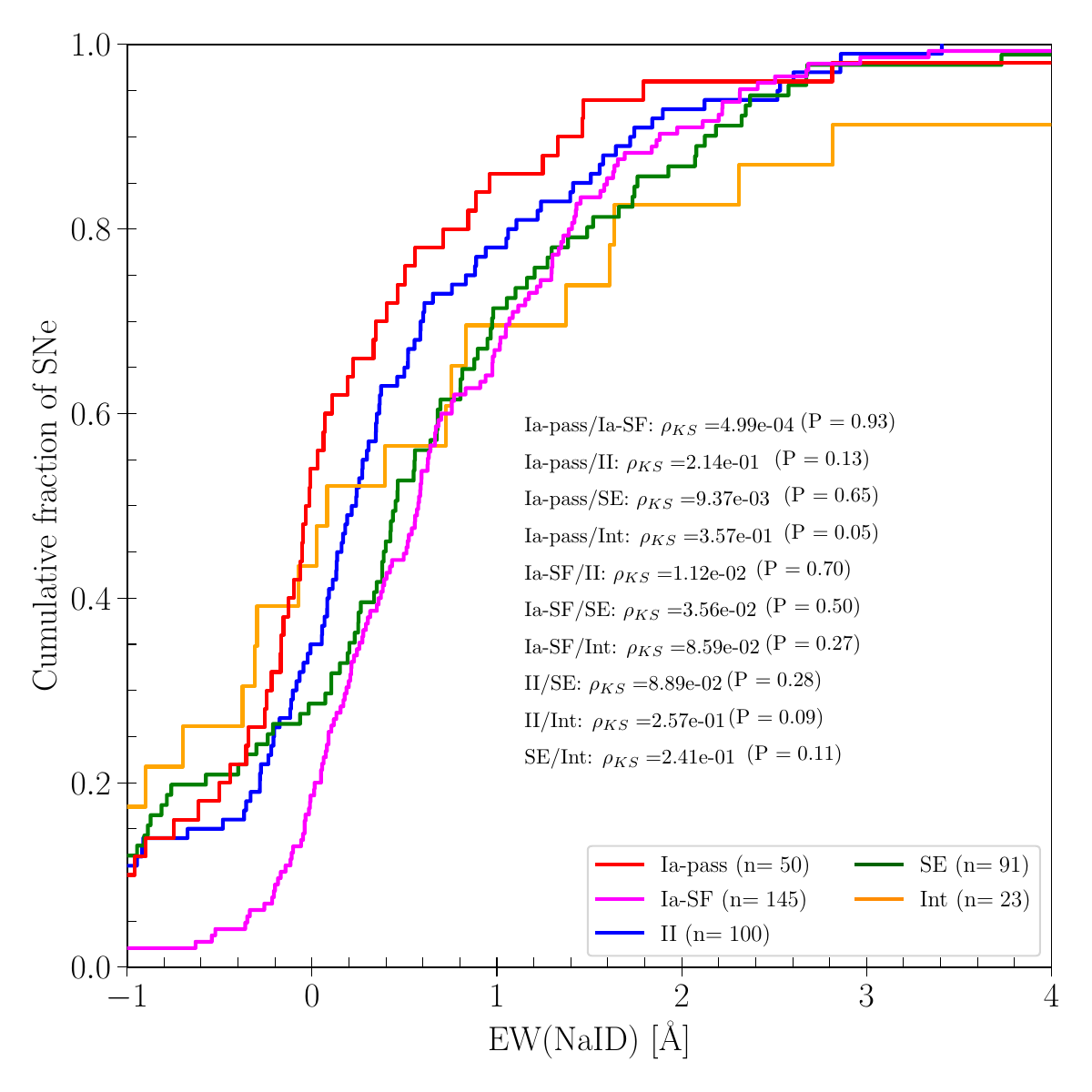}
\includegraphics[width=0.67\columnwidth]{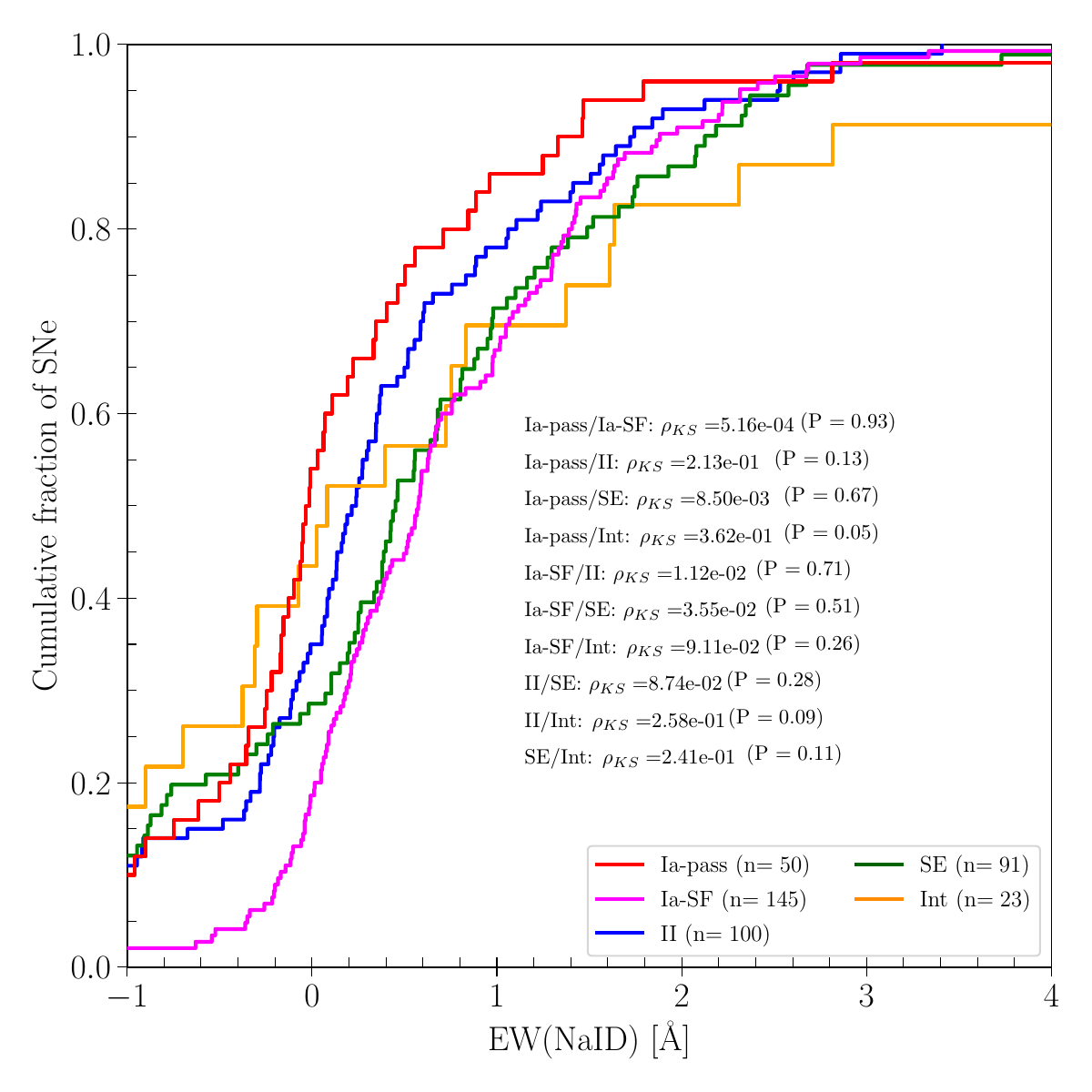}
\caption{Cumulative distributions of the measured \naid\ EW for the full sample \textbf{(left)} and for two independent realisations obtained after applying the random down-sampling procedure (\textbf{middle and right panels)} for each SN type. The number of events of each SN type and the results from the K-S test comparing two different populations are also shown in each panel.
}
\label{fig:downsampling}
\end{figure*}
To evaluate the impact of reducing the number of events in the K-S tests, we applied a random down-sampling procedure to the dataset used in the left panel of Figure~\ref{fig:dist_full}, matching the sample sizes shown in the right panel of the same figure. Specifically, the original samples (Ia-pass$=139$, Ia-SF$=307$, II$=174$, SE$=174$, Int$=85$) were randomly down-sampled to the sizes of the corresponding smaller samples (Ia-pass$=50$, Ia-SF$=145$, II$=100$, SE$=91$, Int$=23$). For each SN type, objects were selected at random and without repetition. This approach preserves the overall shape of the original distributions while mitigating potential biases caused by reducing sample sizes. To account for the intrinsic randomness of the down-sampling procedure, we repeated the process using a Monte Carlo approach with 1000 realisations. The resulting distributions are shown in Figure~\ref{fig:downsampling}, together with the original results (left panel). As shown in the figure, the differences between SN types are generally observed after down-sampling. In particular, the p-values systematically increase while the corresponding probabilities P$_{MC}$ decrease, indicating a reduced statistical significance when controlling for sample size effects. Table~\ref{table:downsampling} presents the p-values and P$_{MC}$ obtained from the original distributions, two representative down-sampling realisations, and the median values derived from the 1000 Monte Carlo realisations.

\begin{table}
\small
\centering
\caption{K-S statistics comparisons.}
\label{table:downsampling}
\begin{tabular}{c|c|c|c|c}
\hline
\hline
\textbf{Comparison} & \textbf{Values from Figure~\ref{fig:dist_full}} &  \multicolumn{3}{c}{\textbf{Random Down-sampled}} \\
\cline{3-5}
&  &  \textbf{Example 1} & \textbf{Example 2} & \textbf{Median} \\ 
\hline
Ia-pass/Ia-SF & 1.3E-07 (100\%) & 5.0E-04 (93\%) & 5.2E-04 (93\%) & 1.4e-03 (87\%) \\
Ia-pass/II    & 1.2E-02 (66\%)  & 2.1E-01 (13\%) & 2.1E-01 (13\%) & 2.0e-01 (14\%) \\
Ia-pass/SE    & 1.4E-08 (100\%) & 9.4E-03 (65\%) & 8.5E-03 (67\%) & 1.2e-03 (89\%) \\
Ia-pass/Int   & 3.7e-02 (53\%)  & 3.6E-01 (5\%)  & 3.6E-01 (5\%)  & 2.0e-01 (13\%) \\
Ia-SF/II      & 1.1e-03 (93\%)  & 1.1E-02 (70\%) & 1.1E-02 (71\%) & 1.5e-02 (64\%) \\
Ia-SF/SE      & 2.1e-01 (9\%)   & 3.6E-02 (50\%) & 3.6E-02 (51\%) & 4.2e-01 (11\%) \\
Ia-SF/Int     & 5.4e-03 (79\%)  & 8.6E-02 (27\%) & 9.1E-02 (26\%) & 4.2e-02 (30\%) \\
II/SE         & 8.5e-05 (99\%)  & 8.9E-02 (28\%) & 8.7E-02 (28\%) & 5.6e-03 (75\%) \\
II/Int        & 7.6e-02 (31\%)  & 2.6E-01 (9\%)  & 2.6E-01 (9\%)  & 2.8e-01 (8\%)  \\
SE/Int        & 9.5e-03 (68\%)  & 2.4E-01 (11\%) & 2.4E-01 (11\%) & 1.8e-01 (15\%) \\
\hline
\end{tabular}
\tablefoot{K-S statistical comparisons of the EW distributions for the original samples and for two distributions derived from the down-sampling procedure. The last column reports the median p-values and the P$_{MC}$ of 1000 realisations.} \end{table}

\section{Supplementary Tables}
\label{ap:tables}

\input{table_ks-dib}
\input{table_median_dib}

\input{table_ks-Ia-pass}
\input{table_ks-Ia-sf}
\input{table_ks-II}
\input{table_ks-SE}
\input{table_ks-Int}


\label{lastpage}

\end{document}

%% file: table_ks_gen.tex
\renewcommand{\arraystretch}{1.3}
\setlength{\tabcolsep}{2.0pt}
\begin{table*}
\fontsize{7.5}{10}\selectfont
\centering
\caption{K-S statistics for various properties divided according to SN type.}
\label{table:KSgen}
\begin{tabular}{c|c|c|c|c|c|c|c|c|c|c}
Property & Ia-pass/Ia-SF & Ia-pass/II & Ia-pass/SE & Ia-pass/Int & Ia-SF/II & Ia-SF/SE & Ia-SF/Int & II/SE & II/Int & SE/Int \\
\hline                                         
\hline
\multicolumn{11}{c}{\textbf{\naid}}\\
\hdashline    
EW          & 1.4E$-7$ (100\%) & 1.2E$-2$ (67\%) & 1.5E$-8$ (100\%) & 3.9E$-2$ (51\%) & 1.1E$-3$ (92\%) & 2.1E$-1$ (10\%) & 5.3E$-3$ (78\%) & 8.6E$-5$ (99\%) & 7.4E$-2$ (31\%) & 9.6E$-3$ (68\%) \\ 
VEL         & 2.5E$-1$ (10\%)  & 3.6E$-1$ (4\%)  & 3.8E$-2$ (43\%)  & 1.5E$-1$ (16\%) & 1.5E$-1$ (15\%) & 9.6E$-2$ (26\%) & 6.1E$-3$ (79\%) & 4.0E$-1$ (5\%)  & 6.3E$-2$ (32\%) & 1.7E$-2$ (59\%) \\
\hdashline  
EW$^{\star}$& 6.4E$-7$ (100\%) & 4.8E$-5$ (99\%) & 1.8E$-5$ (100\%) & 1.1E$-1$ (22\%) & 7.4E$-2$ (30\%) & 1.4E$-1$ (18\%) & 6.1E$-2$ (35\%) & 3.0E$-1$ (7\%)  & 1.7E$-1$ (16\%) & 1.4E$-1$ (19\%) \\ 
\hline
\multicolumn{11}{c}{\textbf{General properties}}\\
\hdashline    
$\overline{\Delta\alpha}$  & 1.8E$-2$ (60\%) & 3.2E$-1$ (55\%) & 4.0E$-3$ (86\%) & 2.0E$-1$ (13\%) & 4.4E$-2$ (39\%) & 4.1E$-1$ (3\%)  & 4.5E$-1$ (4\%) & 4.0E$-2$ (42\%) & 2.7E$-1$ (11\%) & 4.4E$-1$ (4\%) \\  
$i(^{\circ})$              & 6.5E$-4$ (96\%) & 8.0E$-3$ (78\%) & 8.8E$-4$ (92\%) & 8.9E$-2$ (27\%) & 1.3E$-1$ (20\%) & 1.2E$-1$ (22\%) & 3.0E$-1$ (9\%) & 5.2E$-1$ (2\%)  & 4.4E$-1$ (4\%)  & 2.7E$-1$(10\%) \\   
\hline
\multicolumn{11}{c}{\textbf{Local properties}}\\
\hdashline    
M$_*^L$              & 2.3E$-2$ (52\%) & 2.8E$-3$ (88\%) & 1.5E$-3$ (87\%) & 3.1E$-2$ (49\%) & 5.2E$-2$ (37\%) & 7.8E$-2$ (30\%) & 2.6E$-1$ (10\%) & 4.0E$-1$ (5\%)  & 2.6E$-1$ (10\%) & 2.4E$-1$ (11\%) \\  
SFR$_0^L$            & 4.1E$-1$ (4\%)  & 1.4E$-1$ (18\%) & 4.7E$-1$ (3\%)  & 3.5E$-1$ (7\%)  & 2.6E$-1$ (6\%)  & 3.2E$-1$ (5\%)  & 4.3E$-1$ (4\%)  & 2.4E$-2$ (72\%) & 3.1E$-1$ (7\%)  & 3.6E$-1$ (6\%)  \\
sSFR$_0^L$           & 1.5E$-1$ (18\%) & 2.5E$-1$ (10\%) & 7.0E$-2$ (31\%) & 1.2E$-1$ (21\%) & 4.0E$-1$ (2\%)  & 4.4E$-2$ (40\%) & 2.8E$-1$ (10\%) & 2.1E$-2$ (69\%) & 2.1E$-1$ (12\%) & 4.6E$-1$ (4\%)  \\
$t_{\mathrm{age}}^L$ & 2.0E$-2$ (55\%) & 4.3E$-2$ (37\%) & 4.0E$-4$ (94\%) & 1.2E$-2$ (63\%) & 2.6E$-1$ (7\%)  & 1.2E$-2$ (69\%) & 2.1E$-1$ (14\%) & 1.8E$-4$ (99\%) & 1.3E$-1$ (19\%) & 3.4E$-1$ (7\%)  \\
$\tau^L$             & 5.5E$-2$ (40\%) & 5.6E$-2$ (38\%) & 3.6E$-2$ (54\%) & 1.4E$-1$ (18\%) & 3.6E$-2$ (53\%) & 2.6E$-1$ (7\%)  & 2.3E$-1$ (12\%) & 2.0E$-3$ (91\%) & 3.6E$-2$ (41\%) & 2.7E$-1$ (10\%) \\
$A_V^L$              & 3.4E$-1$ (5\%)  & 2.0E$-2$ (54\%) & 2.2E$-1$ (13\%) & 2.6E$-1$ (9\%)  & 1.0E$-3$ (96\%) & 1.7E$-1$ (15\%) & 3.5E$-1$ (7\%)  & 7.5E$-3$ (78\%) & 1.8E$-1$ (15\%) & 4.3E$-1$ (4\%)  \\
$n^L$                & 1.3E$-1$ (19\%) & 2.7E$-2$ (9\%)  & 1.5E$-2$ (18\%) & 2.4E$-1$ (10\%) & 4.2E$-1$ (3\%)  & 5.9E$-1$ (1\%)  & 3.4E$-1$ (6\%)  & 2.6E$-1$ (9\%)  & 3.9E$-1$ (5\%)  & 3.2E$-1$ (8\%)  \\ 
$Z_*^L$              & 1.2E$-2$ (64\%) & 5.4E$-5$ (98\%) & 1.8E$-6$ (100\%)& 4.8E$-4$ (95\%) & 6.6E$-3$ (82\%) & 1.1E$-4$ (99\%) & 6.4E$-2$ (31\%) & 1.1E$-1$ (21\%) & 3.4E$-1$ (7\%)  & 4.3E$-1$ (4\%)  \\
\hdashline  
$sA_V^L\dagger$      & 3.9E$-2$ (41\%) & 2.4E$-3$ (86\%) & 1.8E$-3$ (88\%) & 1.5E$-1$ (17\%) & 2.7E$-1$ (7\%)  & 2.5E$-1$ (10\%) & 3.9E$-1$ (5\%)  & 2.9E$-1$ (8\%)  & 2.9E$-1$ (8\%)  & 2.3E$-1$ (12\%) \\
\hline
\multicolumn{11}{c}{\textbf{Global properties}}\\
\hdashline    
M$_*^G$              & 5.9E$-2$ (37\%) & 7.9E$-7$ (100\%)& 7.3E$-5$ (99\%) & 1.8E$-1$ (15\%) & 1.3E$-3$ (93\%) & 3.0E$-3$ (88\%) & 4.0E$-1$ (5\%)  & 4.0E$-1$ (5\%)  & 1.3E$-1$ (20\%) & 1.7E$-1$ (17\%) \\  
SFR$_0^G$            & 1.3E$-3$ (91\%) & 4.4E$-5$ (100\%)& 8.8E$-4$ (95\%) & 1.4E$-2$ (66\%) & 4.8E$-3$ (79\%) & 2.0E$-1$ (14\%) & 3.3E$-1$ (8\%)  & 9.2E$-2$ (26\%) & 3.6E$-1$ (6\%)  & 4.1E$-1$ (4\%)  \\
sSFR$_0^G$           & 1.2E$-2$ (70\%) & 2.1E$-2$ (57\%) & 3.2E$-2$ (49\%) & 9.0E$-2$ (27\%) & 4.6E$-1$ (4\%)  & 3.6E$-1$ (4\%)  & 3.7E$-1$ (6\%)  & 3.1E$-1$ (8\%)  & 3.4E$-1$ (6\%)  & 3.1E$-1$ (7\%)  \\
$t_{\mathrm{age}}^G$ & 1.4E$-1$ (19\%) & 1.5E$-2$ (79\%) & 5.5E$-2$ (42\%) & 9.8E$-2$ (26\%) & 9.5E$-2$ (25\%) & 1.2E$-1$ (21\%) & 2.9E$-1$ (9\%)  & 3.7E$-1$ (7\%)  & 9.1E$-1$ (25\%) & 2.3E$-1$ (13\%) \\
$\tau^G$             & 4.0E$-2$ (48\%) & 1.9E$-2$ (64\%) & 2.1E$-2$ (62\%) & 1.1E$-1$ (22\%) & 3.2E$-1$ (6\%)  & 2.1E$-1$ (10\%) & 3.3E$-1$ (7\%)  & 5.6E$-2$ (35\%) & 3.5E$-1$ (6\%)  & 2.7E$-1$ (10\%) \\
$A_V^G$              & 1.0E$-1$ (25\%) & 2.3E$-2$ (58\%) & 5.2E$-3$ (81\%) & 8.7E$-1$ (28\%) & 4.1E$-1$ (5\%)  & 1.0E$-1$ (24\%) & 2.5E$-1$ (10\%) & 2.7E$-1$ (7\%)  & 2.5E$-1$ (10\%) & 3.8E$-1$ (5\%)  \\
$n^G$                & 2.5E$-5$ (99\%) & 4.2E$-5$ (99\%) & 4.6E$-6$ (99\%) & 7.8E$-4$ (91\%) & 4.1E$-1$ (4\%)  & 2.3E$-1$ (9\%)  & 3.6E$-1$ (6\%)  & 1.9E$-1$ (13\%) & 3.9E$-1$ (5\%)  & 3.5E$-1$ (6\%)  \\ 
$Z_*^G$              & 8.9E$-2$ (27\%) & 2.4E$-4$ (99\%) & 1.5E$-2$ (65\%) & 1.4E$-1$ (19\%) & 1.9E$-2$ (62\%) & 2.4E$-1$ (9\%)  & 3.6E$-1$ (6\%)  & 1.1E$-1$ (23\%) & 2.6E$-1$ (10\%) & 3.8E$-1$ (6\%)  \\
\hdashline  
$sA_V^G\dagger$      & 4.1E$-2$ (40\%) & 3.8E$-3$ (82\%) & 2.3E$-4$ (96\%) & 3.7E$-1$ (5\%)  & 3.0E$-1$ (8\%)  & 5.9E$-2$ (32\%) & 4.0E$-1$ (6\%)  & 4.3E$-1$ (4\%)  & 3.3E$-1$ (8\%)  & 2.9E$-1$ (10\%) \\
\end{tabular}
\begin{tablenotes}
\small
\item The null hypothesis of the K-S test is that the parameter of two samples divided by the SN type comes from the same parent population. The $p$-value and the probability $\mathbf{P}$ of the $p$-value being lower than 0.05 according to a bootstrap \citepalias[see][]{GG25} are shown. 
\item $^{\star}$ EW values obtained after applying a normalisation by the galaxy offset, SFR$_0^L$ and M$_*^L$.
\item $\dagger$ Specific attenuation, $sA_V=A_V/M_*$. 
\end{tablenotes}
\end{table*}

%% file: table_median.tex
\begin{table}
\centering
\caption{Statistics for the EW and VEL of \naid for different SN types.}
\label{table:median}
\renewcommand{\arraystretch}{1.4}
\begin{tabular}{c|cccc}
\hline
\hline
Type & Nr &  Median & Average & MAD \\ 
\hline
\multicolumn{5} {c} {\naid EW (\AA)}  \\
\hline
Ia-SF &  307 & 0.52 & 0.59 & 0.53 \\
Ia-pass & 139 & $-0.03$ & 0.19 & 0.37 \\
II & 174 & 0.23 & 0.44 & 0.43 \\
SE & 174 & 0.55 & 0.75 & 0.58 \\
Int & 85 & 0.15 & 0.90 & 0.85 \\
\hline
 \multicolumn{5} {c} {\naid VEL (km/s)}  \\
\hline
Ia-SF &  211 & 5 & 0 & 85 \\
Ia-pass & 74 & $-18$ & $-30$ & 87 \\
II & 109 & 20 & $-39$ & 115 \\
SE & 140 & 33 & $-22$ & 104 \\
Int & 70 & $-59$ & $-45$ & 181 \\
\hline
\end{tabular}
\tablefoot{Median, weighted average and median absolute deviation (MAD) of \naid EW and VEL for different SN types.}
\end{table}

%% file: table_ks-dib.tex
\renewcommand{\arraystretch}{1.3}
\setlength{\tabcolsep}{2.0pt}
\begin{table*}
\fontsize{7.5}{10}\selectfont
\centering
\caption{K-S statistics for EW of several absorption lines divided according to SN type.}
\label{table:KSgen-dib}
\begin{tabular}{c|c|c|c|c|c|c|c|c|c|c}
Line     & Ia-pass/Ia-SF   & Ia-pass/II      & Ia-pass/SE      & Ia-pass/Int     & Ia-SF/II        & Ia-SF/SE        & Ia-SF/Int       & II/SE           & II/Int          & SE/Int          \\
\hline                              
\naid    & 1.4E$-7$ (100\%)& 1.2E$-2$ (67\%) & 1.5E$-8$ (100\%)& 3.9E$-2$ (51\%) & 1.1E$-3$ (92\%) & 2.1E$-1$ (10\%) & 5.3E$-3$ (78\%) & 8.6E$-5$ (99\%) & 7.4E$-2$ (31\%) & 9.6E$-2$ (68\%) \\ 
\caii H  & 9.9E$-3$ (72\%) & 7.0E$-2$ (33\%) & 2.2E$-2$ (69\%) & 3.2E$-2$ (58\%) & 1.1E$-2$ (70\%) & 1.6E$-3$ (90\%) & 3.6E$-3$ (83\%) & 3.7E$-1$ (4\%)  & 4.9E$-1$ (3\%)  & 2.1E$-1$ (12\%) \\ 
\caii K  & 4.7E$-4$ (94\%) & 6.2E$-4$ (96\%) & 1.4E$-6$ (100\%)& 4.8E$-2$ (37\%) & 1.4E$-1$ (18\%) & 1.1E$-2$ (68\%) & 3.6E$-1$ (5\%)  & 5.5E$-2$ (38\%) & 3.1E$-1$ (8\%)  & 1.8E$-1$ (15\%) \\ 
\ki 1    & 1.4E$-1$ (19\%) & 1.3E$-1$ (19\%) & 3.0E$-1$ (6\%)  & 4.5E$-2$ (37\%) & 1.0E$-1$ (24\%) & 7.9E$-2$ (31\%) & 1.4E$-1$ (20\%) & 4.0E$-1$ (4\%)  & 3.6E$-1$ (5\%)  & 2.8E$-1$ (8\%)  \\ 
\ki 2    & 1.6E$-1$ (17\%) & 2.5E$-1$ (11\%) & 3.4E$-1$ (7\%)  & 3.5E$-1$ (7\%)  & 4.0E$-3$ (85\%) & 2.2E$-2$ (62\%) & 2.7E$-1$ (8\%)  & 4.6E$-1$ (2\%)  & 1.3E$-1$ (19\%) & 2.1E$-1$ (14\%) \\ 
DIB-4428 & 2.4E$-1$ (12\%) & 2.9E$-1$ (7\%)  & 4.9E$-1$ (3\%)  & 3.5E$-1$ (6\%)  & 4.2E$-3$ (83\%) & 2.7E$-1$ (9\%)  & 6.1E$-4$ (96\%) & 1.6E$-1$ (16\%) & 4.0E$-1$ (4\%)  & 1.2E$-1$ (20\%) \\ 
DIB-5780 & 5.0E$-2$ (36\%) & 1.4E$-1$ (18\%) & 1.2E$-1$ (21\%) & 8.8E$-2$ (26\%) & 4.3E$-2$ (44\%) & 8.9E$-2$ (28\%) & 3.9E$-1$ (5\%)  & 2.8E$-1$ (7\%)  & 1.3E$-1$ (20\%) & 1.8E$-1$ (14\%) \\ 
DIB-6283 & 1.4E$-1$ (19\%) & 6.9E$-2$ (32\%) & 5.7E$-3$ (80\%) & 7.0E$-7$ (100\%)& 1.1E$-1$ (23\%) & 6.5E$-3$ (78\%) & 2.3E$-4$ (95\%) & 1.7E$-1$ (15\%) & 1.1E$-2$ (66\%) & 3.7E$-2$ (43\%) \\ 
\hline
\end{tabular}
\end{table*}

%% file: table_median_dib.tex
\begin{table*}
\centering
\caption{Median and MAD of the EW of several narrow lines for different SN types.}
\label{table:median-dib}
\renewcommand{\arraystretch}{1.4}
\begin{tabular}{c||c|c|c|c|c|c|c|c|c}
\hline
Type & \naid & \caii H & \caii K & \ki1  & \ki2 & DIB-5780 & DIB-4428 & DIB-6283\\ 
\hline
Ia-SF & $0.52\pm0.52$ & $0.38\pm0.42$ & 0.30$\pm$0.37  & $-0.01\pm0.33$ & $-0.03\pm0.28$ & $0.12\pm0.26$ & $0.25\pm0.36$ & $-0.05\pm0.24$\\
Ia-pass & $-0.03\pm0.37$ & $0.07\pm0.36$  & $0.05\pm0.36$ &  $-0.09\pm0.42$ & $-0.04\pm0.39$ & $0.06\pm0.26$ & $0.10\pm0.22$ & $-0.12\pm0.23$\\
II & $0.23\pm0.43$ & $0.14\pm0.56$ & $0.30\pm0.53$  & $0.04\pm0.42$ & $-0.15\pm0.44$ & $0.05\pm0.28$ & $0.03\pm0.20$ & $-0.02\pm0.34$\\
SE & $0.55\pm0.58$ & $0.19\pm0.65$ & $0.50\pm0.59$ & $0.04\pm0.51$ & $-0.14\pm0.50$ & $0.11\pm0.35$ & $0.11\pm0.26$ & $0.04\pm0.39$\\
Int & $0.15\pm0.85$ & $0.08\pm0.71$ & $0.25\pm0.53$  & $0.09\pm0.41$ & $-0.02\pm0.42$ & $0.15\pm0.24$ & $0.02\pm0.19$ & $0.16\pm0.29$ \\
\hline
\end{tabular}

\end{table*}

%% file: table_ks-Ia-pass.tex
\renewcommand{\arraystretch}{1.3}
\begin{table*}
\begin{threeparttable}
\small
\centering
\caption{K-S statistics and correlations for \naid EW and VEL in \textbf{SNe~Ia-pass} divided according to galaxy properties}
\label{table:KS-Ia-pass}
\renewcommand{\arraystretch}{1.5}
\setlength{\tabcolsep}{3.5pt}
\begin{tabular}{c|ccccc|ccccc}
&  & \multicolumn{4} {c} {\bfseries EW} & \multicolumn{5} {|c} {\bfseries VEL}\\
\hline
Property & Nr & $<D^{\mathrm{EW}}_{\mathrm{MC}}>$ & $<p_{MC}^{\mathrm{EW}}>$ & $\mathbf{P}\left(p^{\mathrm{EW}}_{MC}<0.05\right)^{\ast}$ & $r_s^{\mathrm{EW}}$ & Nr & $<D^{\mathrm{vel}}_{\mathrm{MC}}>$ &$<p_{MC}^{\mathrm{vel}}>$ & $\mathbf{P}\left(p^{\mathrm{vel}}_{MC}<0.05\right)^{\ast}$ & $r_s^{\mathrm{vel}}$ \\
\hline                                         
\hline
\multicolumn{10}{c}{\textbf{General properties}}\\
\hdashline 
{\bf T-type$^{\dag}$}                      & 139 & 0.40 & $3.31\times10^{-4}$ & \bf{91\%} & 0.35    & 71 & 0.27 & $2.66\times10^{-1}$ & 10\% & $-0.10$ \\
\boldsymbol{$\overline{\Delta\alpha}$}     & 133 & 0.36 & $1.79\times10^{-3}$ & \bf{91\%} & $-0.30$ & 69 & 0.34 & $9.07\times10^{-2}$ & 25\% & $-0.18$ \\
\boldsymbol{$\Delta\alpha(^{\circ})$}      & 131 & 0.38 & $2.33\times10^{-3}$ & \bf{87\%} & $-0.28$ & 70 & 0.27 & $3.15\times10^{-1}$ & 8\%  & $-0.13$ \\
\boldsymbol{$\Delta\alpha_{\mathrm{DLR}}$} & 133 & 0.35 & $3.00\times10^{-3}$ & \bf{80\%} & $-0.27$ & 69 & 0.40 & $2.21\times10^{-1}$ & 49\% & $-0.19$ \\
$i(^{\circ})$                              & 133 & 0.16 & $4.82\times10^{-1}$ & 3\%       & 0.03    & 69 & 0.32 & $1.42\times10^{-1}$ & 18\% & 0.14    \\
\hline                                                                            
\multicolumn{10}{c}{\textbf{Global properties}}\\
\hdashline                                   
$n^G$                                   & 130 & 0.25 & $7.01\times10^{-2}$ & 30\% & 0.18    & 71 & 0.24 & $4.24\times10^{-1}$ & 5\% & $-0.15$ \\  
$M_*^G$/\Msun                           & 130 & 0.17 & $4.16\times10^{-1}$ & 7\%  & $-0.12$ & 71 & 0.26 & $3.31\times10^{-1}$ & 5\% & $-0.04$ \\
$Z_*^G$/\Zsun                           & 130 & 0.19 & $3.32\times10^{-1}$ & 6\%  & $-0.04$ & 71 & 0.35 & $4.18\times10^{-1}$ & 4\% & 0.09    \\
SFR$_0^G$ (\Msun/yr)                    & 130 & 0.18 & $3.80\times10^{-1}$ & 5\%  & $-0.04$ & 71 & 0.29 & $2.80\times10^{-1}$ & 8\% & $-0.08$ \\
$A_V^G$                                 & 130 & 0.19 & $3.88\times10^{-1}$ & 5\%  & 0.04    & 71 & 0.24 & $5.35\times10^{-1}$ & 3\% & $-0.10$ \\  
$t_{\mathrm{age}}^G$(Gyr)               & 130 & 0.16 & $5.30\times10^{-1}$ & 2\%  & $-0.05$ & 71 & 0.24 & $4.74\times10^{-1}$ & 3\% & 0.06    \\
$\tau^G$(Gyr)                           & 130 & 0.15 & $5.55\times10^{-1}$ & 2\%  & 0.01    & 71 & 0.26 & $3.20\times10^{-1}$ & 6\% & $-0.09$ \\
sSFR$_0^G$ (yr$^{-1}$)                  & 130 & 0.16 & $5.19\times10^{-1}$ & 2\%  & 0.04    & 71 & 0.27 & $3.13\times10^{-1}$ & 6\% & $-0.01$ \\

\hline
\multicolumn{10}{c}{\textbf{Local properties}}\\
\hdashline                             
SFR$_0^L$ (\Msun /yr)                   & 51 & 0.47 & $5.02\times10^{-2}$ & 36\% & 0.25    & 23 & 0.50 & $5.45\times10^{-1}$ & 2\% & 0.18    \\
M$_*^L$/ \Msun                          & 51 & 0.36 & $1.64\times10^{-1}$ & 16\% & 0.26    & 23 & 0.50 & $3.03\times10^{-1}$ & 8\% & 0.41    \\
$\tau^L$(Gyr)                           & 51 & 0.32 & $2.56\times10^{-1}$ & 10\% & $-0.06$ & 23 & 0.47 & $4.08\times10^{-1}$ & 5\% & $-0.10$ \\
sSFR$_0^L$ (yr$^{-1}$)                  & 51 & 0.34 & $2.97\times10^{-1}$ & 9\%  & 0.08    & 23 & 0.50 & $5.38\times10^{-1}$ & 2\% & $-0.07$ \\
$n^L$                                   & 51 & 0.32 & $2.91\times10^{-1}$ & 9\%  & 0.09    & 23 & 0.55 & $3.64\times10^{-1}$ & 5\% & 0.29    \\
$t_{\mathrm{age}}^L$ (Gyr)              & 51 & 0.30 & $3.93\times10^{-1}$ & 5\%  & $-0.09$ & 23 & 0.55 & $3.29\times10^{-1}$ & 6\% & 0.19    \\
$Z_*^L/Z_{\sun}$                        & 51 & 0.27 & $4.89\times10^{-1}$ & 2\%  & $-0.01$ & 23 & 0.47 & $4.60\times10^{-1}$ & 3\% & 0.24    \\
$A_V^L$                                 & 51 & 0.27 & $4.35\times10^{-1}$ & 4\%  & 0.06    & 23 & 0.44 & $4.94\times10^{-1}$ & 3\% & 0.10    \\
\hline  
\end{tabular}
\tablefoot{The null hypothesis probed by the K-S test is that the \naid EW of two samples divided according to a value between the 40\% and 60\% percentile of the galaxy property indicated in the leftmost column of Table~2 in \citetalias{GG25} comes from the same parent population. The K-S statistic, $D$, the $p$-value, the probability $\mathbf{P}$ of the $p$-value being lower than 0.05, and the correlation $r_s$ are shown. Significant rejections of the hypothesis ($\mathbf{P}^{\mathrm{EW}}_{MC}>50\%$) are highlighted in bold. \\
\textbf{Galaxy properties:}
\textbf{\textit{T-type:}} Galaxy classification; \textbf{\textit{$\overline{\Delta\alpha}$:}} SN normalised offset; \textbf{\textit{$\Delta\alpha_{\mathrm{DLR}}$:}} SN directional offset; \textbf{\textit{$\Delta\alpha(^{\circ})$:}} SN angular offset; \textbf{\textit{$i(^{\circ})$:}} Galaxy inclination; \textbf{\textit{M$_*$:}} Stellar mass; \textbf{\textit{SFR:}} Star Formation Rate; \textbf{\textit{sSFR:}} Specific SFR; \textbf{\textit{$Z_*$:}} Stellar metallicity; \textbf{\textit{$t_{\mathrm{age}}$:}} Age; \textbf{\textit{$A_V$:}} Attenuation; \textbf{\textit{$n$:}} dust index; \textbf{\textit{$\tau$:}} e-folding time. 
}
\begin{tablenotes}
\small
\item $^{\ast}$ This probability is obtained from the median of 1000 bootstrap "$z$-matched" simulations on the two samples recalculating the K-S statistic at each iteration and additionally dividing the sample in two at 10 different positions around the median (40-60\% percentile).
\item $^{\dag}$ The T-type samples are divided between spirals and ellipticals (at a fixed T-type $=0$) instead of the range around the median ($<$T-type$>\sim 2$).
\end{tablenotes}
\end{threeparttable}
\end{table*}

%% file: table_ks-Ia-sf.tex
\renewcommand{\arraystretch}{1.3}
\begin{table*}
\begin{threeparttable}
\small
\centering
\caption{K-S statistics and correlations for \naid EW and VEL in \textbf{SNe~Ia-SF} divided according to galaxy properties}
\label{table:KS-Ia-sf}
\renewcommand{\arraystretch}{1.5}
\setlength{\tabcolsep}{3.5pt}
\begin{tabular}{c|ccccc|ccccc}
&  & \multicolumn{4} {c} {\bfseries EW} & \multicolumn{5} {|c} {\bfseries VEL}\\
\hline
Property & Nr & $<D^{\mathrm{EW}}_{\mathrm{MC}}>$ & $<p_{MC}^{\mathrm{EW}}>$ & $\mathbf{P}\left(p^{\mathrm{EW}}_{MC}<0.05\right)^{\ast}$ & $r_s^{\mathrm{EW}}$ & Nr & $<D^{\mathrm{EW}}_{\mathrm{MC}}>$ &$<p_{MC}^{\mathrm{EW}}>$ & $\mathbf{P}\left(p^{\mathrm{EW}}_{MC}<0.05\right)^{\ast}$ & $r_s^{\mathrm{EW}}$ \\
\hline                                         
\hline
\multicolumn{10}{c}{\textbf{General properties}}\\
\hdashline 
\boldsymbol{$\overline{\Delta\alpha}$}       & 299 & 0.40 & $2.42\times10^{-9}$ & \bf{100\%} & $-0.43$ & 205 & 0.15 & $3.03\times10^{-1}$ & 7\%  & 0.02 \\
\boldsymbol{$\Delta\alpha_{\mathrm{DLR}}$}   & 299 & 0.39 & $4.61\times10^{-9}$ & \bf{100\%} & $-0.38$ & 205 & 0.19 & $1.01\times10^{-2}$ & 24\% & 0.05 \\
\boldsymbol{$\Delta\alpha(^{\circ})$}        & 296 & 0.33 & $2.13\times10^{-5}$ & \bf{98\%}  & $-0.23$ & 205 & 0.15 & $3.78\times10^{-1}$ & 4\%  & 0.05 \\
T-type$^{\dag}$                              & 306 & 0.17 & $5.58\times10^{-2}$ & 36\%       & $-0.11$ & 197 & 0.16 & $2.42\times10^{-1}$ & 8\%  & 0.04 \\
$i(^{\circ})$                                & 299 & 0.16 & $8.87\times10^{-2}$ & 27\%       & 0.15    & 205 & 0.20 & $6.80\times10^{-2}$ & 32\% & 0.16 \\
\hline           
\multicolumn{10}{c}{\textbf{Global properties}}\\
\hdashline                                   
$A_V^G$                                      & 279 & 0.15 & $1.44\times10^{-1}$ & 19\% & 0.09    & 197 & 0.15 & $3.40\times10^{-1}$ & 5\%  & $0.04$  \\
$Z_*^G$/\Zsun                                & 279 & 0.13 & $2.68\times10^{-1}$ & 9\%  & $-0.07$ & 197 & 0.15 & $3.52\times10^{-1}$ & 5\%  & $-0.08$ \\
$n^G$                                        & 279 & 0.12 & $4.37\times10^{-1}$ & 4\%  & $-0.08$ & 197 & 0.14 & $4.23\times10^{-1}$ & 4\%  & 0.01    \\
sSFR$_0^G$ (yr$^{-1}$)                       & 279 & 0.11 & $4.44\times10^{-1}$ & 4\%  & 0.02    & 197 & 0.23 & $1.03\times10^{-1}$ & 23\% & $-0.15$ \\
$M_*^G$/\Msun                                & 279 & 0.11 & $4.71\times10^{-1}$ & 3\%  & 0.01    & 197 & 0.14 & $3.98\times10^{-1}$ & 6\%  & 0.04    \\
$\tau^G$(Gyr)                                & 279 & 0.11 & $4.76\times10^{-1}$ & 3\%  & 0.01    & 197 & 0.15 & $1.84\times10^{-1}$ & 15\% & 0.14    \\
SFR$_0^G$ (\Msun/yr)                         & 279 & 0.11 & $5.29\times10^{-1}$ & 2\%  & 0.02    & 197 & 0.19 & $1.44\times10^{-1}$ & 18\% & $-0.14$ \\
$t_{\mathrm{age}}^G$(Gyr)                    & 279 & 0.11 & $5.02\times10^{-1}$ & 2\%  & $-0.01$ & 197 & 0.18 & $1.72\times10^{-1}$ & 16\% & 0.12    \\
\hline
\multicolumn{10}{c}{\textbf{Local properties}}\\
\hdashline                             
\bf{SFR}\boldsymbol{$_0^L$} \bf{(\Msun /yr)} & 149 & 0.43 & $1.84\times10^{-5}$ & \bf{99\%} & 0.48    & 108 & 0.20 & $3.16\times10^{-1}$ & 4\%  & 0.04    \\
\boldsymbol{$A_V^L$}                         & 149 & 0.35 & $1.15\times10^{-3}$ & \bf{91\%} & 0.39    & 108 & 0.20 & $3.54\times10^{-1}$ & 6\%  & $-0.01$ \\
\textbf{M$_*^L$/ \Msun}                      & 149 & 0.34 & $2.22\times10^{-3}$ & \bf{84\%} & 0.38    & 108 & 0.20 & $3.38\times10^{-1}$ & 7\%  & 0.08    \\
\bf{sSFR}\boldsymbol{$_0^L$} \bf{(yr$^{-1}$)}& 149 & 0.32 & $4.64\times10^{-3}$ & \bf{78\%} & 0.32    & 108 & 0.21 & $2.54\times10^{-1}$ & 6\%  & 0.03    \\
\boldsymbol{$t_{\mathrm{age}}^L$} \bf{(Gyr)} & 149 & 0.28 & $1.41\times10^{-2}$ & \bf{58\%} & $-0.25$ & 108 & 0.19 & $4.09\times10^{-1}$ & 4\%  & $-0.03$ \\
$n^L$                                        & 149 & 0.22 & $8.91\times10^{-2}$ & 26\%      & 0.22    & 108 & 0.22 & $2.15\times10^{-1}$ & 13\% & $-0.14$ \\
$Z_*^L/Z_{\sun}$                             & 149 & 0.16 & $4.07\times10^{-1}$ & 3\%       & 0.04    & 108 & 0.18 & $4.68\times10^{-1}$ & 4\%  & 0.05    \\
$\tau^L$(Gyr)                                & 149 & 0.16 & $3.93\times10^{-1}$ & 4\%       & $-0.02$ & 108 & 0.18 & $4.21\times10^{-1}$ & 4\%  & $-0.02$ \\
\hline                                         
\end{tabular}
\tablefoot{Similar to Table~\ref{table:KS-Ia-pass}, but considering SNe~Ia-SF.
}
\end{threeparttable}
\end{table*}

%% file: table_ks-II.tex
\renewcommand{\arraystretch}{1.3}
\begin{table*}
\begin{threeparttable}
\small
\centering
\caption{K-S statistics and correlations for \naid EW and VEL in \textbf{SNe~II} divided according to galaxy properties}
\label{table:KS-II}
\renewcommand{\arraystretch}{1.5}
\setlength{\tabcolsep}{3.5pt}
\begin{tabular}{c|ccccc|ccccc}
&  & \multicolumn{4} {c} {\bfseries EW} & \multicolumn{5} {|c} {\bfseries VEL}\\
\hline
Property & Nr & $<D^{\mathrm{EW}}_{\mathrm{MC}}>$ & $<p_{MC}^{\mathrm{EW}}>$ & $\mathbf{P}\left(p^{\mathrm{EW}}_{MC}<0.05\right)^{\ast}$ & $r_s^{\mathrm{EW}}$ & Nr & $<D^{\mathrm{EW}}_{\mathrm{MC}}>$ &$<p_{MC}^{\mathrm{EW}}>$ & $\mathbf{P}\left(p^{\mathrm{EW}}_{MC}<0.05\right)^{\ast}$ & $r_s^{\mathrm{EW}}$ \\
\hline                                         
\hline
\multicolumn{10}{c}{\textbf{General properties}}\\
\hdashline 
\boldsymbol{$\overline{\Delta\alpha}$} & 151 & 0.32 & $5.57\times10^{-3}$ & \bf{82\%} & $-0.35$ & 97 & 0.22 & $3.18\times10^{-1}$ & 7\% & $-0.13$ \\
$i(^{\circ})$                          & 151 & 0.24 & $6.24\times10^{-2}$ & 34\%      & 0.27    & 97 & 0.21 & $3.35\times10^{-1}$ & 6\% & 0.03    \\
$\Delta\alpha_{\mathrm{DLR}}$          & 150 & 0.31 & $6.46\times10^{-2}$ & 31\%      & $-0.16$ & 98 & 0.19 & $5.12\times10^{-1}$ & 3\% & $-0.06$ \\
$\Delta\alpha(^{\circ})$               & 153 & 0.20 & $1.75\times10^{-1}$ & 15\%      & $-0.11$ & 99 & 0.21 & $3.57\times10^{-1}$ & 7\% & $-0.02$ \\
T-type$^{\dag}$                        & 152 & 0.19 & $2.24\times10^{-1}$ & 12\%      & $-0.14$ & 91 & 0.24 & $3.03\times10^{-1}$ & 8\% & 0.03    \\
\hline                                         
\multicolumn{10}{c}{\textbf{Global properties}}\\
\hdashline                                   
\boldsymbol{$\tau^G$} \bf{(Gyr)}       & 138 & 0.28 & $2.01\times10^{-2}$ & \bf{59\%} & 0.25    & 91 & 0.23 & $2.60\times10^{-1}$ & 10\% & 0.13    \\
sSFR$_0^G$ (yr$^{-1}$)                 & 138 & 0.25 & $5.66\times10^{-2}$ & 31\%      & $-0.18$ & 91 & 0.28 & $1.08\times10^{-1}$ & 22\% & $-0.19$ \\
SFR$_0^G$ (\Msun/yr)                   & 138 & 0.21 & $1.36\times10^{-1}$ & 19\%      & $-0.12$ & 91 & 0.20 & $4.51\times10^{-1}$ & 3\%  & $-0.13$ \\
$t_{\mathrm{age}}^G$(Gyr)              & 138 & 0.21 & $1.73\times10^{-1}$ & 16\%      & 0.10    & 91 & 0.24 & $2.76\times10^{-1}$ & 9\%  & 0.01    \\
$M_*^G$/\Msun                          & 138 & 0.22 & $1.33\times10^{-1}$ & 19\%      & 0.18    & 91 & 0.26 & $1.95\times10^{-1}$ & 14\% & 0.11    \\
$Z_*^G$/\Zsun                          & 138 & 0.16 & $4.39\times10^{-1}$ & 4\%       & $-0.01$ & 91 & 0.21 & $3.71\times10^{-1}$ & 5\%  & $-0.10$ \\
$n^G$                                  & 138 & 0.16 & $4.52\times10^{-1}$ & 4\%       & $-0.04$ & 91 & 0.20 & $4.80\times10^{-1}$ & 3\%  & 0.02    \\
$A_V^G$                                & 138 & 0.17 & $4.31\times10^{-1}$ & 3\%       & 0.03    & 91 & 0.23 & $3.69\times10^{-1}$ & 6\%  & $0.03$  \\
\hline
\multicolumn{10}{c}{\textbf{Local properties}}\\
\hdashline                             
SFR$_0^L$ (\Msun /yr)                  & 103 & 0.25 & $1.27\times10^{-1}$ & 20\%  & 0.18    & 70 & 0.24 & $3.98\times10^{-1}$ & 4\%  & 0.03    \\
M$_*^L$/ \Msun                         & 103 & 0.24 & $1.62\times10^{-1}$ & 16\%  & 0.23    & 70 & 0.24 & $3.63\times10^{-1}$ & 5\%  & 0.12    \\
$A_V^L$                                & 103 & 0.22 & $2.20\times10^{-1}$ & 12\%  & 0.17    & 70 & 0.21 & $5.35\times10^{-1}$ & 2\%  & $-0.01$ \\
sSFR$_0^L$ (yr$^{-1}$)                 & 103 & 0.21 & $3.20\times10^{-1}$ & 8\%   & 0.05    & 70 & 0.24 & $4.42\times10^{-1}$ & 3\%  & 0.02    \\
$Z_*^L/Z_{\sun}$                       & 103 & 0.18 & $4.49\times10^{-1}$ & 4\%   & 0.11    & 70 & 0.27 & $2.23\times10^{-1}$ & 9\%  & $-0.06$ \\
$n^L$                                  & 103 & 0.20 & $3.96\times10^{-1}$ & 4\%   & 0.03    & 70 & 0.25 & $3.46\times10^{-1}$ & 5\%  & 0.01    \\
$t_{\mathrm{age}}^L$ (Gyr)             & 103 & 0.18 & $5.08\times10^{-1}$ & 3\%   & $-0.02$ & 70 & 0.23 & $4.52\times10^{-1}$ & 3\%  & 0.03    \\
$\tau^L$(Gyr)                          & 103 & 0.18 & $4.69\times10^{-1}$ & 3\%   & 0.05    & 70 & 0.23 & $4.14\times10^{-1}$ & 4\% & $-0.02$ \\
\hline                                         
\end{tabular}
\tablefoot{Similar to Table~\ref{table:KS-Ia-pass}, but considering SNe~II.
}
\end{threeparttable}
\end{table*}

%% file: table_ks-SE.tex
\renewcommand{\arraystretch}{1.3}
\begin{table*}
\begin{threeparttable}
\small
\centering
\caption{K-S statistics and correlations for \naid EW and VEL in \textbf{SE-SNe} divided according to galaxy properties}
\label{table:KS-SE}
\renewcommand{\arraystretch}{1.5}
\setlength{\tabcolsep}{3.5pt}
\begin{tabular}{c|ccccc|ccccc}
&  & \multicolumn{4} {c} {\bfseries EW} & \multicolumn{5} {|c} {\bfseries VEL}\\
\hline
Property & Nr & $<D^{\mathrm{EW}}_{\mathrm{MC}}>$ & $<p_{MC}^{\mathrm{EW}}>$ & $\mathbf{P}\left(p^{\mathrm{EW}}_{MC}<0.05\right)^{\ast}$ & $r_s^{\mathrm{EW}}$ & Nr & $<D^{\mathrm{EW}}_{\mathrm{MC}}>$ &$<p_{MC}^{\mathrm{EW}}>$ & $\mathbf{P}\left(p^{\mathrm{EW}}_{MC}<0.05\right)^{\ast}$ & $r_s^{\mathrm{EW}}$ \\
\hline                                         
\hline
\multicolumn{10}{c}{\textbf{General properties}}\\
\hdashline 
\boldsymbol{$\overline{\Delta\alpha}$}    & 143 & 0.31 & $8.07\times10^{-3}$ & \bf{68\%} & $-0.26$ & 114  & 0.18 & $4.32\times10^{-1}$ & 3\%  & $-0.04$ \\
$\Delta\alpha(^{\circ})$                  & 150 & 0.25 & $7.54\times10^{-2}$ & 29\%      & $-0.09$ & 124 & 0.25 & $2.53\times10^{-1}$ & 10\%  & $-0.11$ \\
$\Delta\alpha_{\mathrm{DLR}}$             & 139 & 0.23 & $1.04\times10^{-1}$ & 23\%      & $-0.15$ & 112  & 0.17 & $5.08\times10^{-1}$ & 3\%  & $-0.05$ \\
$i(^{\circ})$                             & 142 & 0.21 & $1.51\times10^{-1}$ & 18\%      & 0.09    & 113  & 0.18 & $4.37\times10^{-1}$ & 4\%  & 0.02    \\
T-type$^{\dag}$                           & 141 & 0.18 & $2.96\times10^{-1}$ & 8\%       & $-0.04$ & 107  & 0.21 & $2.53\times10^{-1}$ & 7\%  & $-0.01$ \\
\hline               
\multicolumn{10}{c}{\textbf{Global properties}}\\
\hdashline                                   
$M_*^G$/\Msun                             & 128 & 0.29 & $3.12\times10^{-2}$ & 43\% & 0.24    & 107 & 0.19 & $4.67\times10^{-1}$ & 4\%  & $-0.01$ \\
$A_V^G$                                   & 128 & 0.22 & $1.91\times10^{-1}$ & 14\% & 0.20    & 107 & 0.21 & $3.68\times10^{-1}$ & 5\%  & $-0.01$ \\
$\tau^G$(Gyr)                             & 128 & 0.21 & $2.22\times10^{-1}$ & 13\% & 0.18    & 107 & 0.27 & $1.10\times10^{-1}$ & 22\% & 0.16    \\
SFR$_0^G$ (\Msun/yr)                      & 128 & 0.21 & $2.05\times10^{-1}$ & 12\% & 0.15    & 107 & 0.22 & $2.67\times10^{-1}$ & 10\% & $-0.16$ \\
$t_{\mathrm{age}}^G$(Gyr)                 & 128 & 0.20 & $2.71\times10^{-1}$ & 10\% & 0.01    & 107 & 0.20 & $3.56\times10^{-1}$ & 6\%  & 0.06    \\
$Z_*^G$/\Zsun                             & 128 & 0.19 & $3.18\times10^{-1}$ & 8\%  & $-0.21$ & 107 & 0.19 & $4.43\times10^{-1}$ & 4\%  & 0.02    \\
$n^G$                                     & 128 & 0.20 & $2.82\times10^{-1}$ & 8\%  & $-0.07$ & 107 & 0.18 & $5.26\times10^{-1}$ & 2\%  & $-0.04$ \\
sSFR$_0^G$ (yr$^{-1}$)                    & 128 & 0.18 & $4.11\times10^{-1}$ & 6\%  & 0.02    & 107 & 0.22 & $2.60\times10^{-1}$ & 11\% & $-0.16$ \\
\hline
\multicolumn{10}{c}{\textbf{Local properties}}\\
\hdashline                             
\boldsymbol{$A_V^L$}                      & 97 & 0.52 & $2.36\times10^{-5}$ & \bf{100\%} & 0.52    & 79 & 0.20 & $5.38\times10^{-1}$ & 2\%  & 0.07 \\
\bf{SFR$_0^L$ (\Msun /yr)}                & 97 & 0.47 & $4.57\times10^{-4}$ & \bf{91\%}  & 0.40    & 79 & 0.25 & $3.10\times10^{-1}$ & 9\%  & 0.17 \\
\bf{M$_*^L$/ \Msun}                       & 97 & 0.41 & $3.10\times10^{-3}$ & \bf{86\%}  & 0.42    & 79 & 0.24 & $3.01\times10^{-1}$ & 9\%  & 0.18 \\
\bf{sSFR$_0^L$ (yr$^{-1}$)}               & 97 & 0.41 & $2.95\times10^{-3}$ & \bf{80\%}  & 0.24    & 79 & 0.23 & $3.63\times10^{-1}$ & 7\%  & 0.09 \\
\boldsymbol{$n^L$}                        & 97 & 0.39 & $4.45\times10^{-3}$ & \bf{78\%}  & 0.31    & 79 & 0.21 & $4.92\times10^{-1}$ & 3\%  & 0.13 \\
$t_{\mathrm{age}}^L$ (Gyr)                & 97 & 0.23 & $2.62\times10^{-1}$ & 10\%       & $-0.12$ & 79 & 0.25 & $3.12\times10^{-1}$ & 7\%  & 0.04 \\
$Z_*^L/Z_{\sun}$                          & 97 & 0.20 & $3.82\times10^{-1}$ & 5\%        & $-0.03$ & 79 & 0.32 & $8.08\times10^{-1}$ & 27\% & 0.25 \\
$\tau^L$(Gyr)                             & 97 & 0.21 & $3.55\times10^{-1}$ & 5\%        & $-0.01$ & 79 & 0.29 & $1.68\times10^{-1}$ & 16\% & 0.11 \\
\hline                                         
\end{tabular}
\tablefoot{Similar to Table~\ref{table:KS-Ia-pass}, but considering SE-SNe.
}
\end{threeparttable}
\end{table*}

%% file: table_ks-Int.tex
\renewcommand{\arraystretch}{1.3}
\begin{table*}
\begin{threeparttable}
\small
\centering
\caption{K-S statistics and correlations for \naid EW and VEL in \textbf{SNe-Int} divided according to galaxy properties}
\label{table:KS-int}
\renewcommand{\arraystretch}{1.5}
\setlength{\tabcolsep}{3.5pt}
\begin{tabular}{c|ccccc|ccccc}
&  & \multicolumn{4} {c} {\bfseries EW} & \multicolumn{5} {|c} {\bfseries VEL}\\
\hline
Property & Nr & $<D^{\mathrm{EW}}_{\mathrm{MC}}>$ & $<p_{MC}^{\mathrm{EW}}>$ & $\mathbf{P}\left(p^{\mathrm{EW}}_{MC}<0.05\right)^{\ast}$ & $r_s^{\mathrm{EW}}$ & Nr & $<D^{\mathrm{EW}}_{\mathrm{MC}}>$ &$<p_{MC}^{\mathrm{EW}}>$ & $\mathbf{P}\left(p^{\mathrm{EW}}_{MC}<0.05\right)^{\ast}$ & $r_s^{\mathrm{EW}}$ \\
\hline                                         
\hline
\multicolumn{10}{c}{\textbf{General properties}}\\
\hdashline 
\boldsymbol{$\overline{\Delta\alpha}$} & 54 & 0.53 & $1.45\times10^{-2}$ & \bf{65\%} & $-0.55$ & 43 & 0.33 & $4.79\times10^{-1}$ & 3\%  & $-0.09$ \\
$\Delta\alpha_{\mathrm{DLR}}$          & 54 & 0.52 & $1.85\times10^{-2}$ & \bf{52\%} & $-0.47$ & 43 & 0.38 & $3.96\times10^{-1}$ & 6\%  & $-0.15$ \\
$i(^{\circ})$                          & 54 & 0.39 & $7.24\times10^{-2}$ & 31\%      & 0.30    & 43 & 0.30 & $5.07\times10^{-1}$ & 3\%  & $-0.07$ \\
$\Delta\alpha(^{\circ})$               & 57 & 0.37 & $1.84\times10^{-1}$ & 14\%      & $-0.32$ & 45 & 0.43 & $2.11\times10^{-1}$ & 14\% & $-0.04$ \\
T-type$^{\dag}$                        & 53 & 0.27 & $5.32\times10^{-1}$ & 2\%       & 0.01    & 32 & 0.27 & $6.35\times10^{-1}$ & 1\%  & $-0.20$ \\
\hline              
\multicolumn{10}{c}{\textbf{Global properties}}\\
\hdashline                                   
$M_*^G$/\Msun                          & 42 & 0.60 & $1.77\times10^{-2}$ & \bf{52\%} & 0.39    & 32 & 0.38 & $4.67\times10^{-1}$ & 4\%  & 0.05    \\
$Z_*^G$/\Zsun                          & 42 & 0.45 & $1.71\times10^{-1}$ & 16\%      & 0.03    & 32 & 0.40 & $3.94\times10^{-1}$ & 5\%  & 0.14    \\
$\tau^G$(Gyr)                          & 42 & 0.44 & $1.59\times10^{-1}$ & 15\%      & 0.18    & 32 & 0.36 & $5.64\times10^{-1}$ & 3\%  & 0.19    \\
$t_{\mathrm{age}}^G$(Gyr)              & 42 & 0.42 & $2.23\times10^{-1}$ & 11\%      & 0.11    & 32 & 0.39 & $5.01\times10^{-1}$ & 3\%  & 0.08    \\
sSFR$_0^G$ (yr$^{-1}$)                 & 42 & 0.33 & $3.72\times10^{-1}$ & 6\%       & $-0.04$ & 32 & 0.43 & $3.37\times10^{-1}$ & 8\%  & 0.34    \\
$A_V^G$                                & 42 & 0.36 & $3.55\times10^{-1}$ & 5\%       & 0.01    & 32 & 0.38 & $5.23\times10^{-1}$ & 3\%  & $-0.20$ \\
$n^G$                                  & 42 & 0.35 & $5.09\times10^{-1}$ & 3\%       & 0.08    & 32 & 0.38 & $4.84\times10^{-1}$ & 3\%  & 0.32    \\
SFR$_0^G$ (\Msun/yr)                   & 42 & 0.33 & $4.78\times10^{-1}$ & 3\%       & 0.25    & 32 & 0.50 & $2.67\times10^{-1}$ & 10\% & 0.21    \\
\hline
\multicolumn{10}{c}{\textbf{Local properties}}\\
\hdashline                             
SFR$_0^L$ (\Msun /yr)                  & 25 & 0.64 & $6.06\times10^{-2}$ & 33\% & 0.53    & 17 & 0.45 & $6.37\times10^{-1}$ & 1\% & 0.03    \\
sSFR$_0^L$ (yr$^{-1}$)                 & 25 & 0.53 & $2.72\times10^{-1}$ & 11\% & 0.38    & 17 & 0.50 & $4.60\times10^{-1}$ & 3\% & $-0.03$ \\
$\tau^L$(Gyr)                          & 25 & 0.48 & $2.84\times10^{-1}$ & 10\% & $-0.37$ & 17 & 0.50 & $4.34\times10^{-1}$ & 4\% & $-0.08$ \\
$A_V^L$                                & 25 & 0.50 & $2.68\times10^{-1}$ & 10\% & 0.38    & 17 & 0.50 & $5.28\times10^{-1}$ & 2\% & $-0.02$ \\
M$_*^L$/ \Msun                         & 25 & 0.50 & $2.67\times10^{-1}$ & 10\% & 0.40    & 17 & 0.50 & $5.64\times10^{-1}$ & 2\% & 0.05    \\
$t_{\mathrm{age}}^L$ (Gyr)             & 25 & 0.45 & $5.05\times10^{-1}$ & 3\%  & $-0.12$ & 17 & 0.50 & $5.86\times10^{-1}$ & 1\% & $-0.08$ \\
$Z_*^L/Z_{\sun}$                       & 25 & 0.40 & $5.98\times10^{-1}$ & 2\%  & 0.15    & 17 & 0.50 & $5.75\times10^{-1}$ & 2\% & 0.01    \\
$n^L$                                  & 25 & 0.36 & $6.38\times10^{-1}$ & 2\%  & 0.06    & 17 & 0.50 & $6.60\times10^{-1}$ & 1\% & $-0.10$ \\
\hline                                         
\end{tabular}
\tablefoot{Similar to Table~\ref{table:KS-Ia-pass}, but considering SNe-Int.
}
\end{threeparttable}
\end{table*}